%% file: main.tex
\documentclass[aps,prr,reprint,twocolumn,superscriptaddress,floatfix,nofootinbib,amsmath, amssymb, longbibliography]{revtex4-1}

\usepackage[utf8]{inputenc}
\usepackage{graphicx}
\usepackage{dcolumn}
\usepackage{bm}
\usepackage{color}
\usepackage[normalem]{ulem}

\def\spose#1{\hbox to 0pt{#1\hss}}
\def\lesssim{\mathrel{\spose{\lower 3pt\hbox{$\mathchar"218$}}
 \raise 2.0pt\hbox{$\mathchar"13C$}}}
\def\gtrsim{\mathrel{\spose{\lower 3pt\hbox{$\mathchar"218$}}
 \raise 2.0pt\hbox{$\mathchar"13E$}}}

\begin{document}

\title{Shape and Size Tunability of Sheets of Interlocked Ring Copolymers}

\author{Juan Luengo-M\'arquez}
\email{juan.luengo@uam.es}
\affiliation{Departamento de Física Teórica de la Materia Condensada, Universidad Autónoma de Madrid, 28049 Madrid (Spain)}
\affiliation{Instituto Nicol\'as Cabrera, Universidad Autónoma de Madrid, 28049 Madrid (Spain)}

\author{Salvatore Assenza}
\affiliation{Departamento de Física Teórica de la Materia Condensada, Universidad Autónoma de Madrid, 28049 Madrid (Spain)}
\affiliation{Instituto Nicol\'as Cabrera, Universidad Autónoma de Madrid, 28049 Madrid (Spain)}
\affiliation{Condensed Matter Physics Center (IFIMAC), Universidad Autónoma de Madrid, 28049 Madrid (Spain)}

\author{Cristian Micheletti}
\email{cristian.micheletti@sissa.it}
\affiliation{Scuola Internazionale Superiore di Studi Avanzati (SISSA), Via Bonomea 265, I-34136 Trieste, Italy.}

\graphicspath{{../Figures/}}

\begin{abstract}
{\bf Abstract.}
\input{sections/abstract.tex}
\end{abstract}
\date{\today}
\maketitle

\input{sections/introduction.tex}
\input{sections/methods.tex}
\input{sections/results_discussion.tex}
\input{sections/conclusions.tex}

\section*{Acknowledgements}
JLM would like to acknowledge the Erasmus+ Programme for Traineeships for financial support.
The project that gave rise to these results received the support
of a fellowship from the “la Caixa” Foundation (ID
100010434) and from the European Union’s Horizon research
and innovation programme under Marie Skłodowska-Curie
grant agreement no. 847648. The fellowship code is LCF/BQ/
PI20/11760019. C.M. acknowledges support from MUR grant PRIN-2022R8YXMR and PNRR grant CN\_00000013\_CN-HPC, M4C2I1.4, spoke 7, funded by NextGenerationEU.
S.A. acknowledges support from a Ramón y Cajal Fellowship (ref. RYC2022-037744-I), funded by MICIU/AEI/10.13039/501100011033 and FSE+.
We are grateful to Alexander Klotz for helpful feedback on the manuscript and for sharing in return his manuscript titled "Chirality Effects in Molecular Chainmail" co-authored with Michael Dimitriyev and Caleb Anderson that is being jointly uploaded on arXiv.

\renewcommand\refname{References}


\bibliography{bbl_SISSA}

\phantom{XXXX}\\
\phantom{XXXX}\\
\phantom{XXXX}\\
\hrule
\phantom{XXXX}\\

\setcounter{figure}{0}
\renewcommand{\figurename}{Fig.}
\renewcommand{\thefigure}{S\arabic{figure}}
\setcounter{section}{0}
\renewcommand{\thesection}{S\arabic{section}}
\setcounter{equation}{0}
\renewcommand{\theequation}{S\arabic{equation}}

\noindent\LARGE{\textbf{Supplementary Information}

\begin{figure*}[ht!]
        \centering
\includegraphics[width=0.65\textwidth,keepaspectratio]{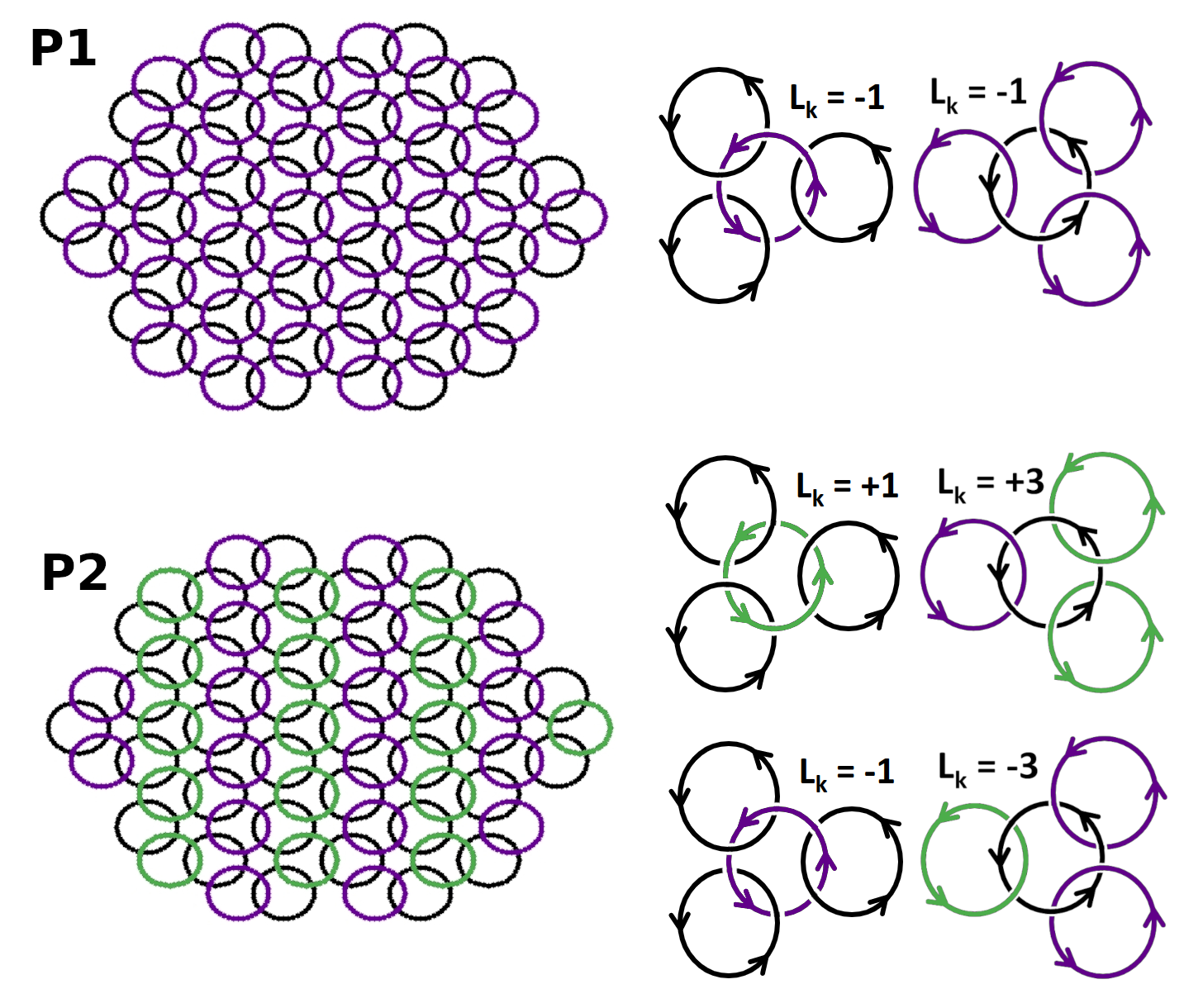}
        \caption{{\bf Linking modes in $P1$ and $P2$ lattices.} Shape and linking pattern of the catenated membranes with $n=68$ rings. Linking number associated to each of the linking modes of every ring with three neighbors.}
\label{fig:connectivity_All}
\end{figure*}

\begin{figure}[ht!]
        \centering
\includegraphics[width=0.45\textwidth,keepaspectratio]{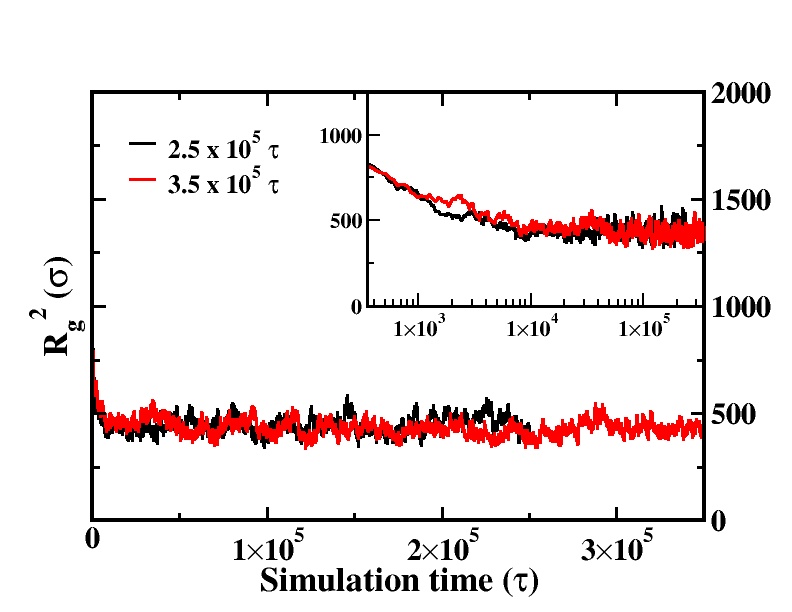}
        \caption{{\bf Characteristic relaxation times.} For each composition and linking pattern, we first run a $2.5  \times 10^{5} \tau$ simulation. We then test the convergence of the measured properties running a simulation of $3.5  \times 10^{5} \tau$. Here we show as an example the global $R_{g}^{2}$ of the fully rigid composition with $P1$. The inset displays the logarithmic scale. The relaxation time is roughly $5  \times 10^{4} \tau$, which is the time that we discarded for the analysis.
        }
\label{fig:relaxation}
\end{figure}

\begin{figure}
        \centering
\includegraphics[width=0.45\textwidth,keepaspectratio]{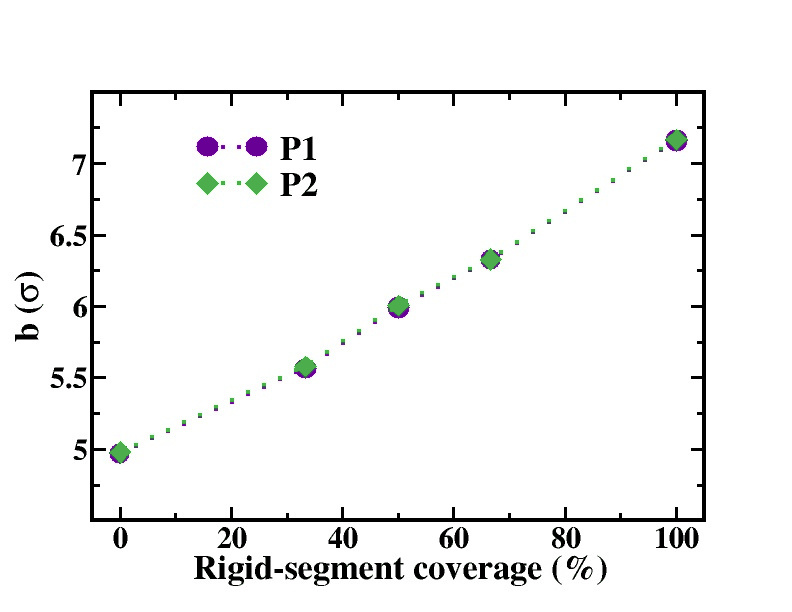}
        \caption{{\bf Average mechanical bond length.} Average mechanical bond length for both linking patterns: distance between centers of linked rings. Error bars are present, although smaller than the symbols. The errors are computed as the estimate of the standard error of the sample mean considering the average values from the simulation of $2.5  \times 10^{5} \tau$ and the one of $3.5  \times 10^{5} \tau$.
        }
\label{fig:distcencen}
\end{figure}
\begin{figure}
        \centering
\includegraphics[width=0.49\textwidth,keepaspectratio]{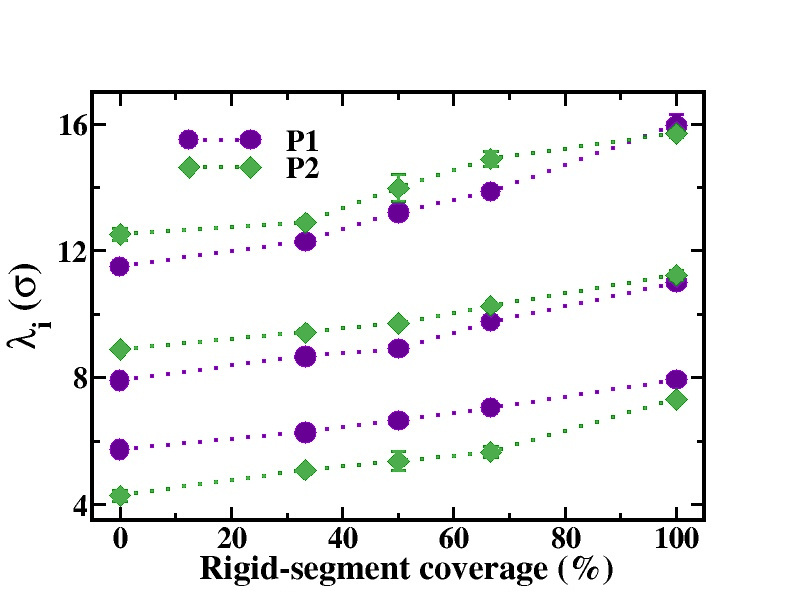}
        \caption{{\bf Shape of membranes.} Eigenvalues of the gyration tensor of the whole membranes ($\lambda_{1}^{2} > \lambda_{2}^{2} > \lambda_{3}^{2}$) without normalization ($\sqrt{<\lambda_{i}^{2}>}$). Error bars are computed as the estimate of the standard error of the sample mean considering the average values from the simulation of $2.5  \times 10^{5} \tau$ and the one of $3.5  \times 10^{5} \tau$.
        }
\label{fig:shapeMem}
\end{figure}
\begin{figure}
        \centering
\includegraphics[width=0.49\textwidth,keepaspectratio]{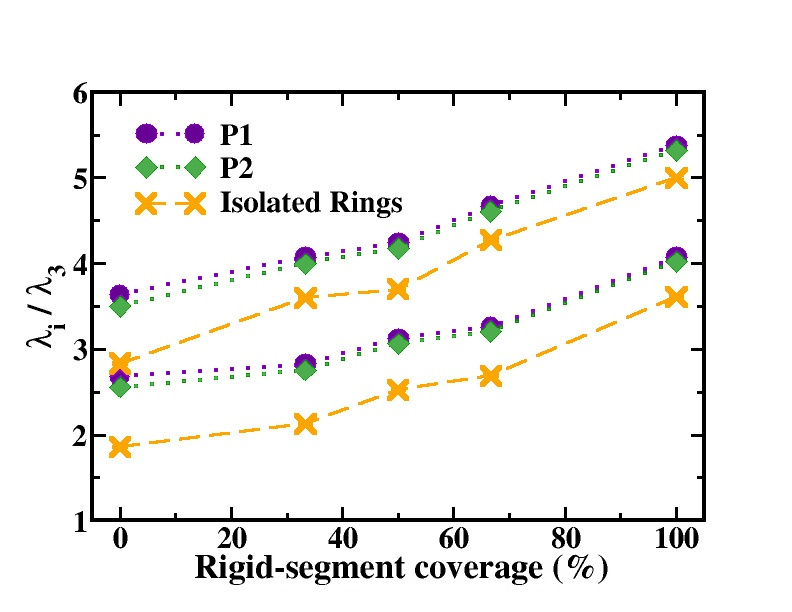}
        \caption{{\bf Shape of single rings.} $\sqrt{<\lambda_{i}^{2}>/<\lambda_{3}^{2}>}$ for $i=1,2$ of single rings. Rings within membranes with $P1$ (violet dots) and with $P2$ (green diamonds) do not differ significantly, as it is generally the case for local properties. Both seem stretched when compared to isolated rings (orange crosses), whose $\lambda_{1}$ and $\lambda_{2}$ are closer to $\lambda_{3}$. As expected, this stretching effect is less important in the case of fully rigid rings. Error bars are present, although smaller than the symbols. The errors in $P1$ and $P2$ are computed as the estimate of the standard error of the sample mean considering the average values from the simulation of $2.5  \times 10^{5} \tau$ and the one of $3.5  \times 10^{5} \tau$. The errors for the isolated rings are computed with block analysis and bootstrapping.        }
\label{fig:shapeSR}
\end{figure}

\section{Model for contacts among particles}
\begin{small}
Here we show a detailed derivation of Eq. 7 in the main text. For any given pair of interlocked rings, we consider the different costs in conformational energy $\Delta F_{ij}$ associated with any combination of contacts, using as a reference each isolated ring

\begin{equation}
        \Delta F_{ff} \equiv \Delta F_{f} + \Delta F_{f}\ ,
\notag
\end{equation}
\begin{equation}
        \Delta F_{fr} \equiv \Delta F_{f} + \Delta F_{r}\ ,
\notag
\end{equation}
\begin{equation}
        \Delta F_{rr} \equiv \Delta F_{r} + \Delta F_{r}\ ,
\label{eq:free_energies}
\end{equation}
\noindent where $\Delta F_{i}$ is the change in conformational free energy of a ring whenever it exposes a particle of type $i$ as a contact with another ring. Therefore, in Eq. \ref{eq:free_energies}, we have assumed that the free energy of a contact is a simple addition of free-energy changes associated with each of the fragments establishing the contact, independently of the nature of the partner fragment.

We can then define the excess free energy $\Delta F \equiv \Delta F_{f} - \Delta F_{r}$ as the net free-energy change associated with a contact established by a \textit{flexible} particle as compared to a \textit{rigid} one. We also define $x$ as the fraction of rigid particles over the total number of particles per ring ($0\leq x\leq 1$).

Then the probability of each type of contact is proportional to the probability of a random sampling (assuming that the number of particles is large) times a Boltzmann factor accounting for the statistical weight of each sampling

\begin{equation}
        P_{ff}(x) \propto (1-x)^{2}\exp\left(-\frac{\Delta F_{f-f}}{k_{B}T}\right) = (1-x)^{2}\mu^{2}\omega^{2} ,
\notag
\end{equation}
\begin{equation}
        P_{fr}(x) \propto 2x(1-x)\exp\left(-\frac{\Delta F_{f-r}}{k_{B}T}\right) = 2x(1-x)\mu^{2}\omega ,
\notag
\end{equation}
\begin{equation}
        P_{rr}(x) \propto x^{2}\exp\left(-\frac{\Delta F_{r-r}}{k_{B}T}\right) = x^{2}\mu^{2},
\label{eq:boltzmann_factors}
\end{equation}
where the last expression of each appears after substituting the corresponding free energy change with its definition in Eq. \ref{eq:free_energies} and then applying the definitions
\begin{equation}
        \mu \equiv \exp\left(-\frac{\Delta F_{r}}{k_{B}T}\right), \hspace{0.5cm} \omega \equiv \exp\left(-\frac{\Delta F}{k_{B}T}\right).
\label{eq:last_definitions}
\end{equation}

Eq. 7 in the main text follows after normalizing all probabilities in Eq. \ref{eq:boltzmann_factors} with $P_{ff}(x)+P_{fr}(x)+P_{rr}(x) = \mu^{2}\left[(1-x)^{2}\omega^{2}+2x(1-x)\omega+x^{2}\right]$.
\end{small}

\begin{figure*}[ht!]
        \centering
\includegraphics[width=1.1\textwidth,keepaspectratio]{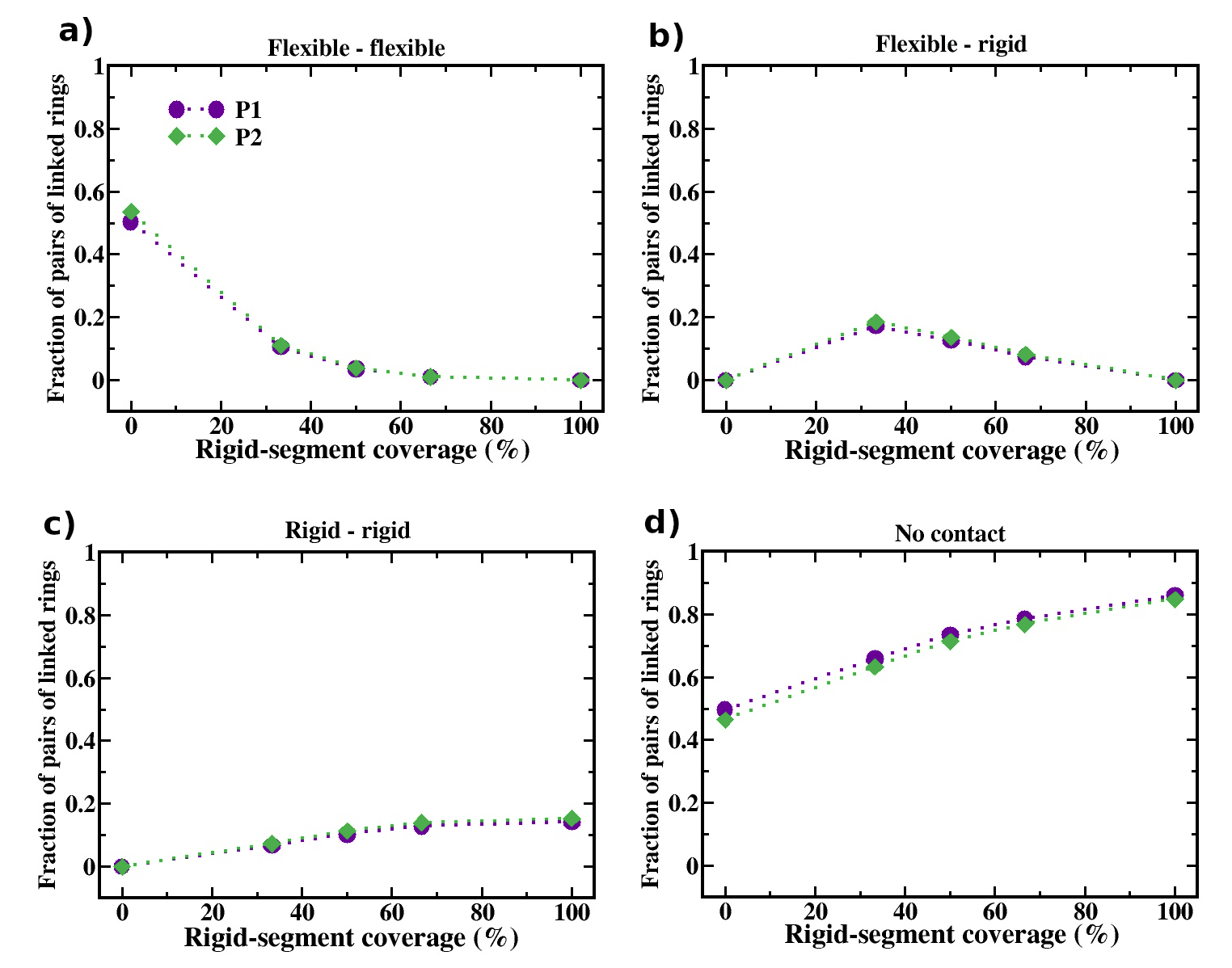}
        \caption{{\bf Contacts between ring pairs.} Fraction of ring pairs having a) flexible-flexible contact, b) flexible-rigid contact, c) rigid-rigid contact, and d) no contact. For each pair of linked rings, we identify the pair of particles --- belonging to each of the rings --- i) whose distance is smaller than $\sqrt[6]{2}\sigma$, and ii) which are closer to each other. We assign the type of contact depending on the particle type. If no pair of particles for a given ring pair fulfills the conditions i) and ii), we assign "no contact". The errors are computed as the estimate of the standard error of the sample mean considering the average values from the simulation of $2.5  \times 10^{5} \tau$ and the one of $3.5  \times 10^{5} \tau$.}
\label{fig:contactsRingsPairs}
\end{figure*}

\begin{figure*}[ht!]
        \centering
\includegraphics[width=0.9\textwidth,keepaspectratio]{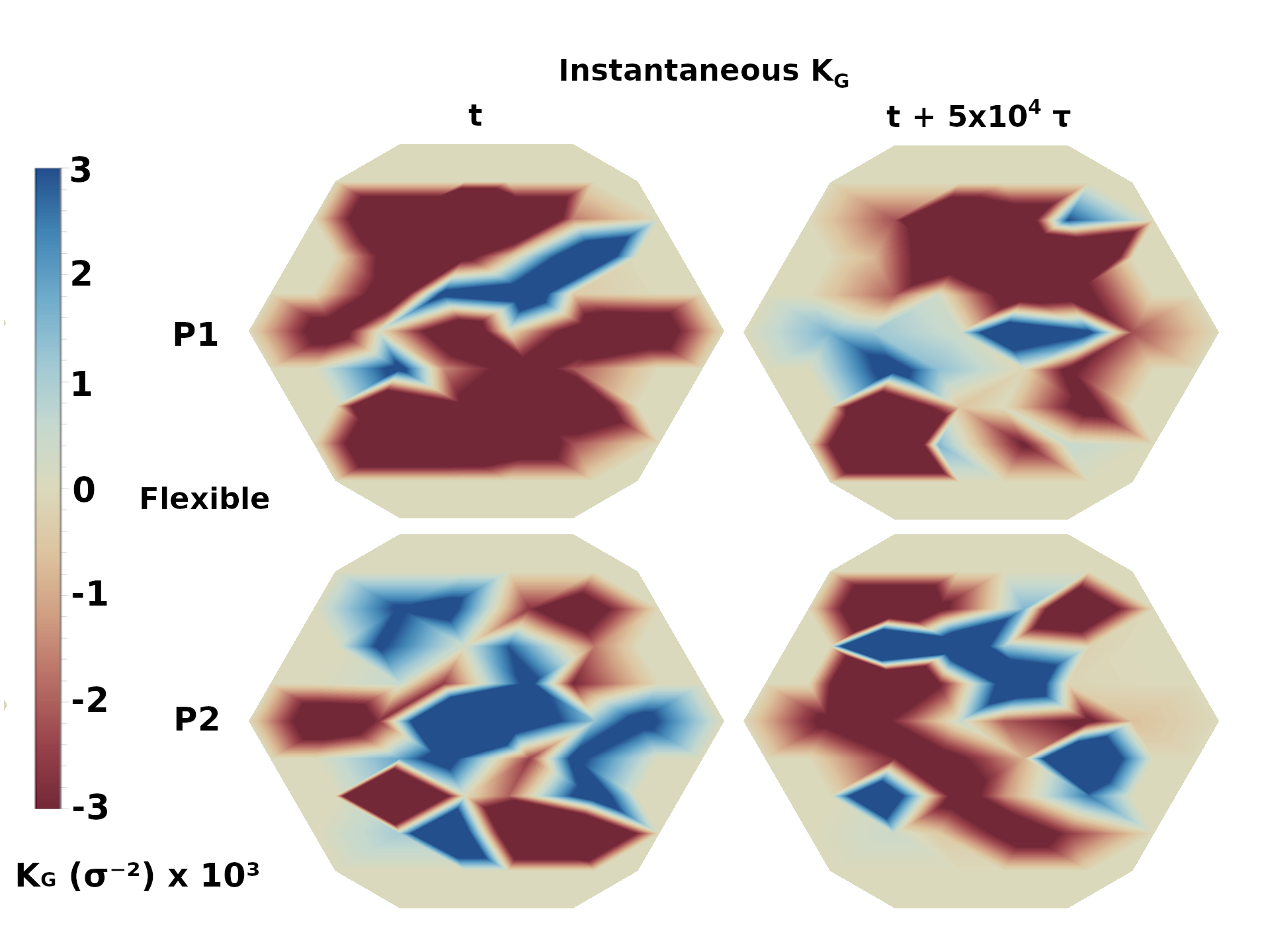}
        \caption{{\bf Instantaneous $K_G$ heatmaps.} Instantaneous $K_{G}$ heatmaps using the $2.5\times 10^{5}\tau$ simulation of the fully-flexible composition. Top line corresponds to P1 chainmails, while bottom line are P2 chainmails. For each linking pattern, we show the $K_G$ per vertex of two frameshots separated a time $5\times 10^{5}\tau$, similar to the typical autocorrelation time of the gyration radius of the chainmails. The legend bar is saturated for better visualization.
        }
\label{fig:instantaneousKG}
\end{figure*}

\begin{figure}[ht!]
        \centering
\includegraphics[width=0.62\textwidth,height=1.0\textwidth,keepaspectratio]{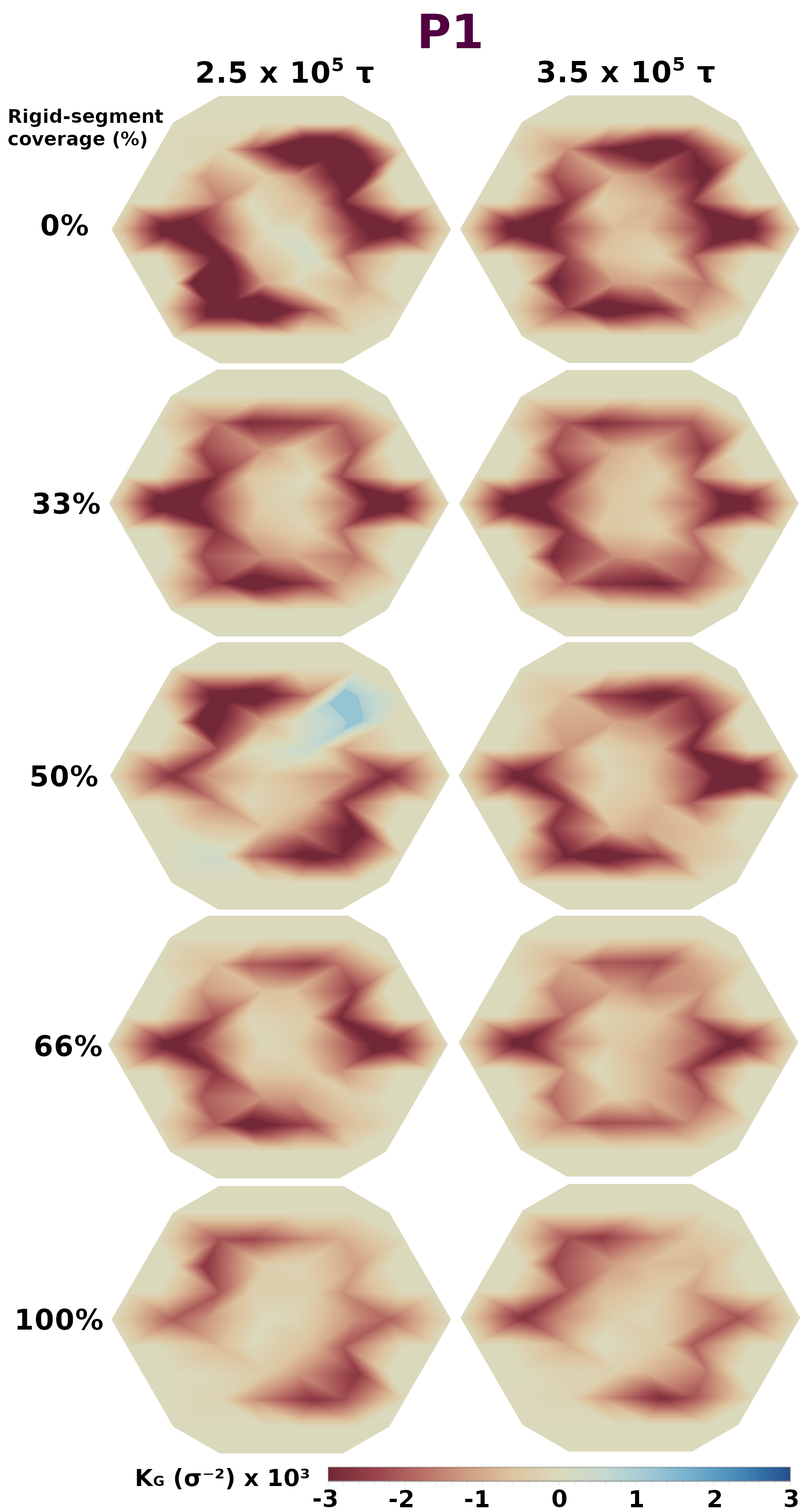}
        \caption{{\bf P1 chainmail, Gaussian curvature at each vertex.} $K_{G}$ per vertex for $P1$ membranes at all compositions. Left column displays the $2.5\times 10^{5}\tau$ simulation and right column the $3.5\times 10^{5}\tau$ simulation. In the $50 \%$ case, the line with more negative values of $K_{G}$ has different inclinations, pointing out a spontaneous symmetry breaking. The legend bar is saturated in the negative side for better visualization.        }
\label{fig:P1AllKg}
\end{figure}

\begin{figure}[ht!]
        \centering
\includegraphics[width=0.62\textwidth,height=1.0\textwidth,keepaspectratio]{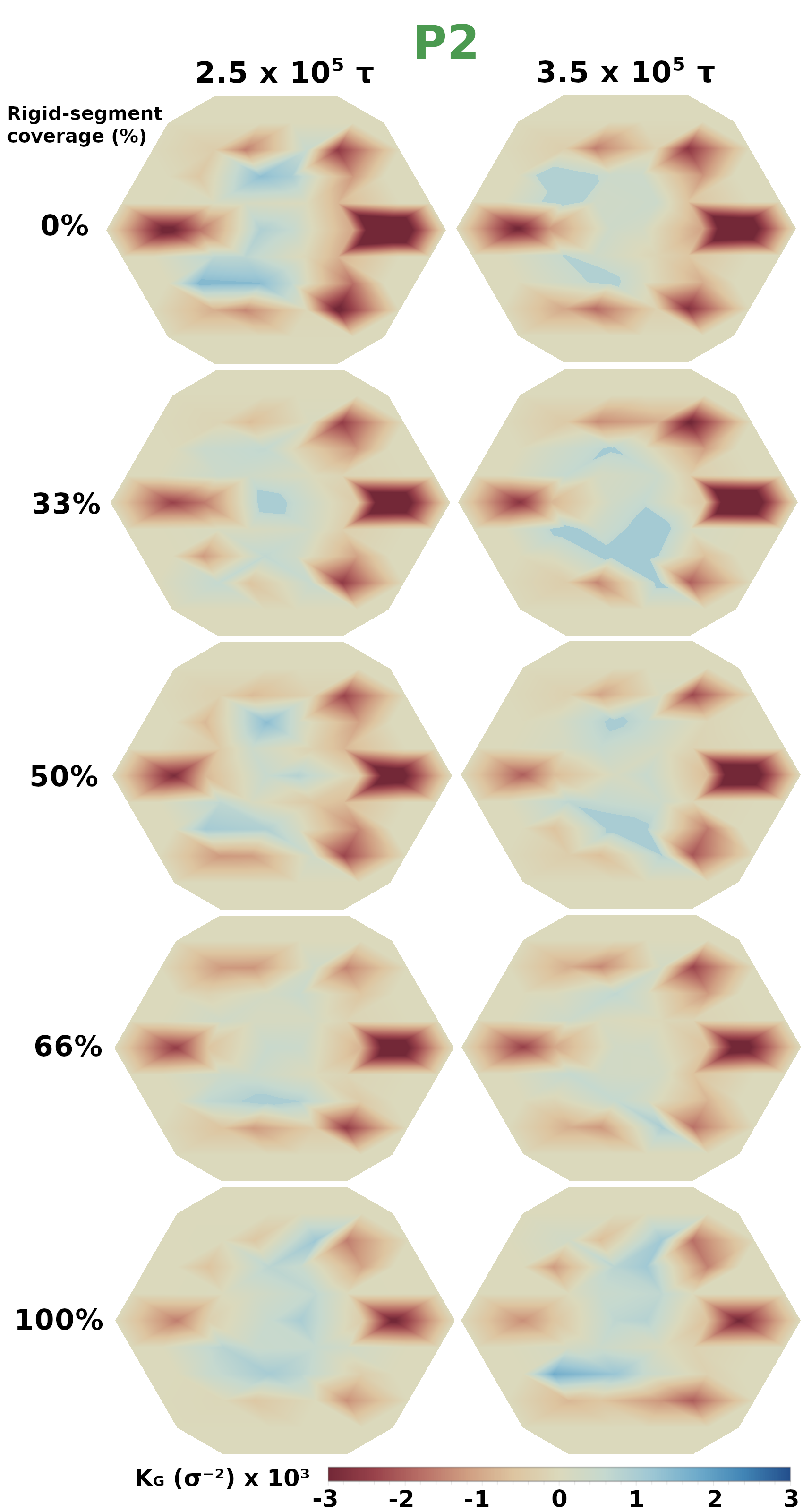}
        \caption{{\bf P2 chainmail, Gaussian curvature at each vertex.} $K_{G}$ per vertex for $P2$ membranes at all compositions. Left column displays the $2.5\times 10^{5}\tau$ simulation and right column the $3.5\times 10^{5}\tau$ simulation. The legend bar is saturated in the negative side for better visualization.        }
\label{fig:P2AllKg}
\end{figure}

\begin{figure}[ht!]
        \centering
\includegraphics[width=0.49\textwidth]{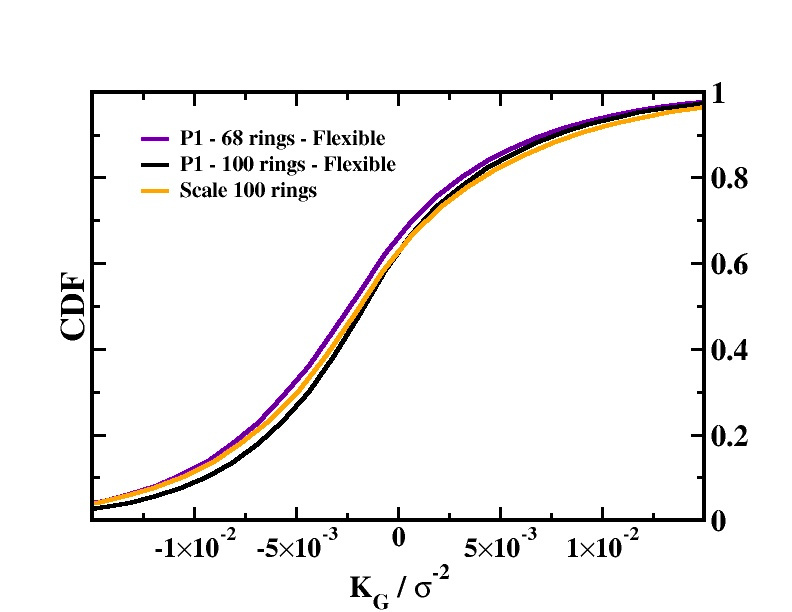}
        \caption{{\bf Scaling of membranes with number of rings.} Curvature scales with membrane size. Cumulative Density Functions (CDFs) of $K_{G}$ for fully flexible $P1$ membranes with 68 rings (violet dots) and 100 rings (black line). Orange line represents the re-scaled distribution of the black line with $q = 1.06 \pm 0.04$. The scaling factor was computed minimizing the quadratic error of the re-scaled PDF relative to the PDF of the membrane with 68 rings, and its error span the values of $q$ that provide a quadratic error under $5 \%$ relative to the peak of the reference distribution.
        }
\label{fig:scaling_Size}
\end{figure}

\end{document}

%% file: sections/abstract.tex
Mechanically bonded membranes of interlocked ring polymers are a significant generalization of conventional elastic sheets, where connectivity is provided by covalent bonding, and represent a promising class of topological meta-materials. In this context, two open questions regard the large-scale reverberations of the heterogeneous composition of the rings and the inequivalent modes of interlocking neighboring rings. We address these questions with Langevin dynamics simulations of chainmails with honeycomb-lattice connectivity, where the rings are block copolymers with two segments of different rigidity. We considered various combinations of the relative lengths of the two segments and the patterns of the over- and under-passes linking neighboring rings. We find that varying ring composition and linking patterns have independent and complementary effects. While the former sets the overall size of the chainmail, the latter defines the shape, enabling the selection of starkly different conformation types. Notably, one of the considered linking patterns favors saddle-shaped membranes, providing a first example of spontaneous negative Gaussian curvature in mechanically bonded sheets. The results help establish the extent to which mechanically bonded membranes can differ from conventional elastic membranes, particularly for the achievable shape and size tunability.

%% file: sections/introduction.tex
\section{Introduction}
The large-scale conformations of polymerized or crystalline membranes, elastic sheets formed by particles tethered or bonded in a two-dimensional network, can be highly susceptible to local structural details\cite{kantor1986statistical,kantor1993excluded,barsky1994molecular,bowick2001statistical,witten2007stress,sakamoto2009two,gandikota2023crumpling}. For instance, introducing isolated defects can cause otherwise flat membranes to buckle out-of-plane and acquire positive or negative Gaussian curvatures over large scales\cite{seung1988defects,gompper1997freezing,bowick2001statistical,zhang2014defects}. More strikingly, introducing even a short-ranged self-avoidance in tethered membranes can prevent their transition from flat to crumpled as temperature increases\cite{plischke1988absence,abraham1989molecular,murat1996molecular,kantor1993excluded,spector1994conformations}. These consequential demonstrations of the coupling of small and large-scale structural properties have no analog in linear polymers and originate from the unique balance between entropic and enthalpic (elastic) contributions to the free-energy of quasi-two-dimensional systems embedded in three-dimensions. Devising ways to offset this balance has been a major objective both theoretically and for applicative purposes, especially the design of tunable materials \cite{zhang2015mechanically,yllanes2017thermal,fokker2019crumpling}.

In addition to the above endeavors, entirely new perspectives for designing membranes and tuning their elastic properties are being opened by recent advancements in supramolecular chemistry, which have made it possible to obtain extended low-dimensional materials by harnessing mechanical or topological bonds in place of conventional covalent ones\cite{NiuCatenaneReview2009,Gil-RamirezCatenaneReview2015,RowanCatenaneScience2017,HartMITReview2021,orlandini2021topological,bai2023oligo}
. Examples include the synthesis or assembly of responsive molecular constructs \cite{ErbasArtificial2015,CollierCatenaneSwitch2000},  networks of interlocked molecules \cite{lee2012mechanical,SpakowitzOlympic2018,august2020self,SakaiTopogami2021}, long linear catenanes \cite{RowanCatenaneScience2017,SougataToroidalCatenane2020}. In consideration of their one-dimensional character, the latter systems have represented the simplest setting for understanding the effects of mechanical bonding on conformational\cite{DehaghaniCatenaneExponent2020,LeiCatenaneShape2021,LiDoubleScaling2021,chiarantoni2022rigidity}, dynamical\cite{RauscherIsolated2018,RauscherThermoMelt2020,RausherDynamicsMelt2020,farimani2024effects} and mechanical\cite{RowanCatenaneScience2017,Caraglio_et_al_polymers_2017} properties, also by contrast with conventional bonding. It is thus natural to ask what novel large-scale behavior can emerge in two-dimensional elastic sheets that, instead of being made by bonded particles, consist of a chainmail of linked molecular rings.

Polymerized and mechanically bonded membranes are similar in some respects and radically different in others, making their comparison ideal to further our understanding of elastic sheets and topological materials. A fundamental common element is that the underlying topology is permanently fixed for both types of membranes, insofar as conventional and mechanical bonds are "unbreakable". In addition, the distance of mechanically bonded rings can fluctuate within specific limits, similar to the distance of neighboring nodes in tethered sheets. However, while traditional tethered membranes are constituted by (linear) polymer chains converging and binding at the nodes, in chainmails the (closed) polymer chains serve at one time as tethers as well as nodes, their bonding arising from topological constraints.

These differences are expected to be consequential for at least two reasons. First, large-scale properties of membranes can be sensitive to small-scale features, as noted before. Second, topological constraints can endow mechanically bonded structures with emerging physical properties. A case in point are circular catenanes of rigid rings, which behave similarly to elastic ribbons, even though the rings interact only sterically and not via bending or torsional potentials\cite{tubiana2022}. In addition, the fact that the elementary units of concatenated structures are extended structures, e.g.~ring polymers, generally suffices to introduce new physical regimes not accessible to the covalently bonded counterparts, from dynamical relaxation\cite{RauscherIsolated2018,RauscherThermoMelt2020,RausherDynamicsMelt2020} to the response to mechanical stretching\cite{RowanCatenaneScience2017,Caraglio_et_al_polymers_2017,Caraglio_et_al_macromol_2017} or spatial confinement \cite{chiarantoni2023}, to the available modes of entanglement\cite{dehaghani2023topological}.

Further reasons for considering sheets of linked ring polymers come from biological systems, as best illustrated by the mitochondrial DNA of trypanosomes\cite{fairlamb1978,ChenKinetoplastNet1995} that consists of thousands of interlocked DNA rings. Intriguingly, the rings show a significant presence of sequence stretches known as A-tracts\cite{johnson2013poly}, which are characterized by a stiff mechanical response\cite{assenza2022accurate}, thus providing an example of rings with heterogeneous elasticity in vivo.
Single-molecule experiments have recently clarified that these long-known DNA chainmails have the shape of relatively smooth curved membranes\cite{klotz2020,soh2020deformation,he2023}. The membrane is bound by a rigid perimeter, possibly itself formed by redundantly linked DNA rings\cite{ragotskie2024effect}, but is otherwise flexible and hence endowed with significant conformational plasticity that confers unusual mechanical and dynamical properties to the system. For instance, the DNA chainmail can transition reversibly from expanded to collapsed states as the concentration of crowders in solution is varied\cite{yadav2021phase}, but it deforms continuously and without the equivalent of a coil-stretch transition in elongational flows \cite{soh2020deformation}. Various theoretical and computational models have been introduced to understand the observed properties of kinetoplast DNA\cite{arsuaga2014properties,diao2015orientation,ibrahim2018estimating,klotz2020,orlandini2021,polson2021,he2023}, including how they depend on the network of the linked rings\cite{polson2021}.
Notably, the lateral and transverse size of rigid-ring chainmails were found to scale with the system area similarly to flat covalent membranes, and yet, the chainmails invariably featured a spontaneous curvature and precisely a positive Gaussian curvature, absent from conventional membranes\cite{polson2021}. The positive Gaussian curvature was consistently observed across different system sizes and types of linking networks and could thus be ascribed to the topological bonding, which introduces anisotropies in the steric interactions of the rigid rings\cite{polson2021}.

Two general aspects have remained virtually unexplored for mechanically bonded membranes:
(i) the effect of sequence-dependent heterogeneity of the constitutive rings, particularly the sequence-dependent bending rigidity\cite{geggier2010sequence,marin2021,roldan2024systematic}, and (ii) the effects of the inequivalent modes of linking the same set of neighboring rings. Various considerations suggest that both aspects could be consequential for the properties of chainmails. On the one hand, sequence heterogeneity has been previously studied in systems where topological constraints were intra-molecular, namely block-copolymer knotted rings where the different flexibility of the segments was varied explicitly, via the local bending rigidity \cite{orlandini2016local}, or implicitly, via charged/neutral character of the monomers and the ionic strength \cite{tagliabue2020interface,tagliabue2021tunable}.
In both cases, the sequence-modulated elasticity could either pin or delocalize the essential crossings of the knot, allowing, in turn, to tune the global metric properties. Thus, it is relevant to ask whether similar pinning of the essential crossings can be achieved for the inter-molecular topological constraints of chainmail and what the implications are for the conformational properties. On the other hand, the inequivalent modes in which a ring can be linked to its neighbors, that is, the inequivalent patterns of over- and under-passes, arguably provide the most straightforward way of exploring the cooperative effects of mechanical bonding and establishing whether and to what extent mechanically-bonded membranes are affected by features beyond those encoded in the pairwise linking graph of the chainmail.

Here, we study both aspects with Langevin dynamics simulations on block-copolymer chainmails with honeycomb lattice topology and different patterns of linking modes for the constitutive rings, which comprise a rigid and flexible segment. We considered five different compositions of the rings, from fully flexible to rigid, and two different linking patterns. For each of the ten combinations of ring composition and linking pattern, we sampled the equilibrium ensemble of the mechanically bonded membranes and analyzed various local and global observables to characterize how the interplay of local flexibility and linking modes defines the chainmail properties.

The main findings of our study are three. First, we observe that different modes of linking rings to their neighbors can dramatically alter the conformational ensemble of chainmails with the same honeycomb connectivity. Second, we show that a specific pattern of linking modes systematically favors saddle-shaped membranes, thus providing the first example of negative Gaussian curvature. Finally, we find that varying the composition of the block-copolymer rings can alter membrane size and bias the nature of the contacting (interlocked) segments but has no significant effects on membrane shape.

%% file: sections/methods.tex
\section{Methods}

\subsection{Model: chainmail types}

\begin{figure*}[ht!]
        \centering
\includegraphics[width=0.85\textwidth,height=1.01\textwidth,keepaspectratio]{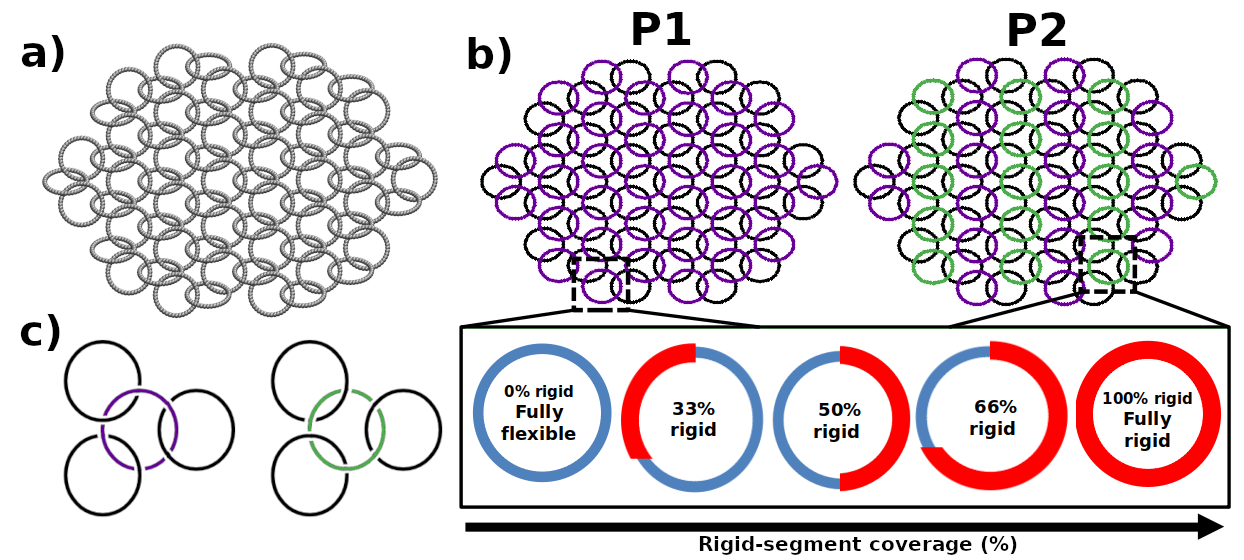}
        \caption{Shape and linking patterns of the considered P1 and P2 chainmails with $n=68$ rings. a) Iinitial configuration of the P1 chainmail. b) Schematic representation of the different patterns of linking modes defining P1 and P2 chainmails; the rings are color-coded based on the succession of over and underpasses  established with their neighbours, as illustrated in panel c). The individual rings are circular di-block copolymers. We considered five different compositions for the di-block rings, corresponding to rigid-segment coverage (percentage of monomers in the rigid segment) of $0\%$, $33\%$, $50\%$, $67\%$, and $100\%$.}
\label{fig:lattice}
\end{figure*}

We considered systems of rings of beads interlocked in a chainmail with the connectivity of a honeycomb lattice consisting of two interleaved triangular sublattices. The mechanical bonding between neighboring rings is provided by the Hopf-type linking, the simplest instance of topological interlocking (Fig.~\ref{fig:lattice}a).
There exist several different modes for interlocking the rings of one triangular sublattice with those of the other, corresponding to different successions of over- and under-passes\cite{tubiana2022}.
To explore the impact of different modes on the conformational properties of the membranes, we consider here two types of chainmails, P1 and P2 in Fig.~\ref{fig:lattice}b. In the case of P1, all rings are connected with the same mode (Fig. S1 top in the Supplementary Material). The linking pattern of P2 was instead
chosen to include multiple linking modes into the same membrane, thus enabling the exploration of the effect of mode heterogeneity (Fig.~S1 bottom in the Supplementary Material).

The initial conformations of P1 and P2 were obtained starting from the same overlapping coplanar honeycomb arrangement of rings of $m=40$ beads of size $\sigma$. All rings were perfectly circular with radius $r\simeq 6.4\sigma$, and the spacing of neighboring rings was set to
approximately $1.5 r$. We retained rings at a distance smaller than about $52.44 \sigma$ from a designated central site, obtaining quasi-hexagonal, hence roughly circular, cutouts of 68 rings.

The first chainmail type, P1, was obtained by tilting the rings in the two triangular sublattices by opposite off-plane rotations. For the system at hand, setting the rotation angle to $+\pi/8$ for the rings in sublattice A (purple) and $-\pi/8$ for sublattice B (black) yielded the sought linking pattern.

The alternative type, P2, was obtained with a different strategy. Rings in sublattice B (black) were kept in their coplanar configurations. Rings in sublattice A were divided into two sets, corresponding to the green and purple alternating columns in the rightmost panel of Fig.~\ref{fig:lattice}b. Next, green and purple rings were deformed with suitable out-of-plane undulations to realize the over- and under-passes sketched in Fig.~\ref{fig:lattice}c.
In the resulting staggered pattern, the over- and under-passes of the sublattice ``A'' rings with their nearest neighbors are reversed in alternating rows.

\subsection{Model: ring polymers}
The excluded-volume interaction of any pairs of monomers in the same or different rings is treated with a standard Weeks-Chandler-Andersen (WCA) potential, corresponding to a truncated and shifted Lennard-Jones potential,
\begin{equation}
U_{\rm WCA} =
\begin{cases}
         4\epsilon\left[\left(\frac{\sigma}{r}\right)^{12}-\left(\frac{\sigma}{r}\right)^{6} \right]+\epsilon & \mbox{if }r\leq \sqrt[6]{2}\sigma,\\
         0 & \mbox{otherwise,}
\end{cases}
\label{eq:lj}
\end{equation}
where the parameter $\epsilon$ sets the energy scale.

The backbone connectivity of the rings was provided by adding a bond potential between consecutive monomers in the form of a standard finitely extensible nonlinear elastic (FENE) term\cite{kremer1990}
\begin{equation}
U_{\rm FENE} = -\frac{K_{\rm FENE} R^{2}_{0}}{2}\ln\left[1-\left(\frac{r}{R_{0}}\right)^{2} \right]\ ,
\label{eq:fene}
\end{equation}
for monomer distances $r\leq R_{0}$ and $U_{FENE} = \infty$ for $r > R_{0}$.
The parameters were set to the standard values $K_{\rm FENE}=30 \ \epsilon/\sigma^{2}$ and $R_{0}=1.5 \ \sigma$\cite{kremer1990}.

We considered diblock copolymer rings with one rigid segment of $n_r=\{0,13,20,27,40\}$ monomers and a flexible segment composed of the remaining $m-n_r$ monomers.
In the present context, a monomer is considered ``rigid'' or ``flexible'' according to the stiffness of the angle centered on it.
The limiting cases $n_r=0$ and $n_r=40$ correspond to fully rigid and fully-flexible rings. The bending potential is

\begin{equation}
        U_{\rm bend} =  \sum_{i=1}^{m} \kappa_{\rm bend}(i)  \, \left(1 - \vec{u}_{i} \cdot \vec{u}_{i+1} )\ , \right .
\label{eq:angular}
\end{equation}
where $\vec{u}_{i}$ is the unit vector from monomer $i-1$ to monomer $i$, assuming periodic bead indexing along the ring contour. The
bending stiffness was set to $\kappa_{\rm bend}(i)=10\epsilon$ and $\kappa_{\rm bend}(i)=0$ for rigid and flexible monomers, respectively.

The five above-mentioned block copolymer compositions that we considered corresponded to 0, 33\%, 50\%, 67\%, and 100\% of particles in a ring belonging to the rigid segment and were adopted uniformly for all rings in the chainmail.

\subsection{Langevin molecular dynamics simulations}
The canonical equilibrium properties of the chainmails were studied with Langevin molecular dynamics simulations, which were integrated with the LAMMPS software package\cite{thompson2022lammps}. We set the temperature of the system by fixing $k_BT=\epsilon$.
We used a periodic cubic simulation box large enough to accommodate the entire initial (planar) chainmail. The dynamics was integrated with a damping factor of $2\tau$
and a time step of $0.005\tau$, where $\tau=\sigma\sqrt{m/\epsilon}$ is the characteristic simulation time and $m$ is the monomer mass, set equal to unity. For each combination of linking pattern and block-copolymer ring composition, we collected two independent trajectories of duration $2.5 \times 10^{5} \tau$ and $3.5 \times 10^{5} \tau$, respectively, discarding an initial interval of $5 \times 10^{4} \tau$ corresponding to the system relaxation time, see Fig. S2 in the Supplementary Material. Configurations were sampled at intervals of $50\tau$. The adequate coverage of the sampling was verified {\em a posteriori} from the consistency of the expectation values of various observables computed separately for the two trajectories.

\subsection{Observables}
To characterize the conformational ensemble of the chainmails we used a combination of local and global observables.

First, for each sampled chainmail configuration, we computed the gyration tensor, $\mathbf{R}$, whose general entry is:
\begin{equation}
        R_{\alpha,\beta} = \frac{1}{N}\sum_{i=1}^{N}(r_{i,\alpha}-r_{CM,\alpha})(r_{i,\beta}-r_{CM,\beta})\ ,
\label{eq:gyration_tensor}
\end{equation}
where $N$ is the total number of monomers in the chainmail, $\alpha$ and $\beta$ run over the three Cartesian components, and $r_{i,\alpha}$ and $r_{CM,\alpha}$ are the $\alpha$ components of the position vectors of the $i$th monomer and of the center of mass of the chainmail, respectively.

The ranked eigenvalues of the gyration tensor, $\lambda_{1}^{2} \geq \lambda_{2}^{2} \geq \lambda_{3}^{2}$, were used to characterize the shape anisotropy of the chainmail and compute its squared gyration radius, $R_{g}^2=\lambda_{1}^{2}+\lambda_{2}^{2}+\lambda_{3}^{2}$. The canonical expectation values of the same quantities were obtained by averaging over the sampled conformations at given chainmail type and ring composition.

\begin{figure}[ht!]
        \centering{
\includegraphics[width=0.40\textwidth,height=0.40\textwidth,keepaspectratio]{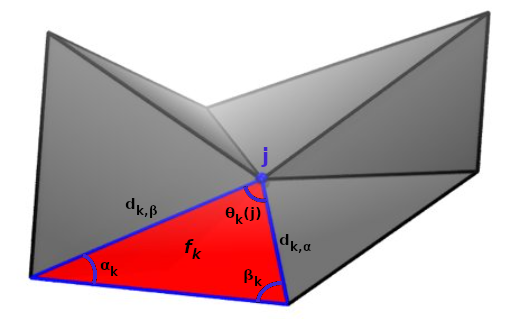}
        \caption{Computation of the local Gaussian curvature. The angle of facet $k$ (red area) at the central vertex $j$ is $\theta_{k}(j)$. The contribution of facet $k$ to $A(j)$ in Eq. \ref{eq:gaussian_curvature} is $\frac{1}{8}(d_{k,\alpha}^{2}\cot{\alpha_{k}} + d_{k,\beta}^{2}\cot{\beta_{k}})$ if the triangle $f_{k}$ is acute or right. Instead, if the triangle $f_{k}$ is obtuse, the contribution of facet $k$ to $A_{j}$ is $A_{k}/2$ if $\theta_{k}$ is obtuse, or $A_{k}/4$ if $\alpha_{k}$ or $\beta_{k}$ are obtuse, where $A_{k}$ is the area of the triangle $f_{k}$.}
\label{fig:KGCalc}}
\end{figure}

We analyzed the spontaneous curvature of the chainmail by computing the local Gaussian curvatures in the neighborhood of each ring using the method of ref.~\cite{meyer2003}. The algorithm applies to three-dimensional embeddings of triangular meshes and thus it is well suited to the honeycomb connectivity of our chainmails.
Specifically, a natural triangulation of the neighborhood of ring $i$ is obtained by connecting its center of mass with those of its nearest neighbor rings of the same ($A$ or $B$) sublattice. This procedure yields six or fewer triangular facets depending on whether the ring $i$ is inside or at the boundary of the chainmail.
Based on the Gauss-Bonnet theorem, the local Gaussian curvature is then computed as:

\begin{equation}
        K_{G}(i) =  \left(2\pi - \sum_{k=1}^{n_f(i)}\theta_{k}(i) \right)\frac{1}{A(i)}\ ,
\label{eq:gaussian_curvature}
\end{equation}
where $n_f(i)$ is the number of triangular facets impinging on ring $i$, $\theta_{k}(i)$ is the vertex angle of facet $k$ at $j$ (see Fig.
\ref{fig:KGCalc}) and $A(i)$ is a surface area term as defined in Fig.~\ref{fig:KGCalc}.

Following ref.~\cite{polson2021}, we computed the global Gaussian curvature of the entire chainmail as
\begin{equation}
        \overline{K}_{G} =  \frac{\sum_{i}K_{G}(i) \, A(i)}{\sum_{i}A(i)}\ ,
\label{eq:avr_gaussian_curvature}
\end{equation}
\vspace{0.10cm}
where $\sum_{i}$ denotes the sum over all rings in the chainmail.

%% file: sections/results_discussion.tex
\section{Results and discussion}

We used Langevin molecular dynamics simulations to explore the conformational properties of honeycomb chainmails of diblock copolymer rings, each made of one rigid and one flexible segment. Our specific aim was to understand whether the large-scale properties of chainmails with the same honeycomb connectivity could significantly depend on the composition of the rings and the inequivalent modes that the rings can be linked with their neighbors.
To this end, we considered various relative sizes of the rigid and flexible segments and two different patterns of chainmail linking, as summarised in Fig.~\ref{fig:lattice}.

For the block copolymer composition, we considered five different rigid-segment coverages, corresponding to 0, 33, 50, 67, and 100\% of ring monomers belonging to the rigid segment.

For the linking patterns, we exploited the bipartite nature of the honeycomb lattice, made of two interleaved triangular sublattices, to design two different chainmail types, as sketched in Fig.~\ref{fig:lattice}b. The first type, P1, is homogeneous in that rings in the two triangular sublattices present the same linking mode, i.e., a succession of over and underpasses, with their mechanically-bonded neighbors, modulo finite size effects. The second type, P2, is instead heterogeneous, with the rings of one triangular sublattice switching between the two linking modes of Fig.~\ref{fig:lattice}c in alternating columns.

\begin{figure*}[ht!]
        \centering
\includegraphics[width=1.05\textwidth,height=1.01\textwidth,keepaspectratio]{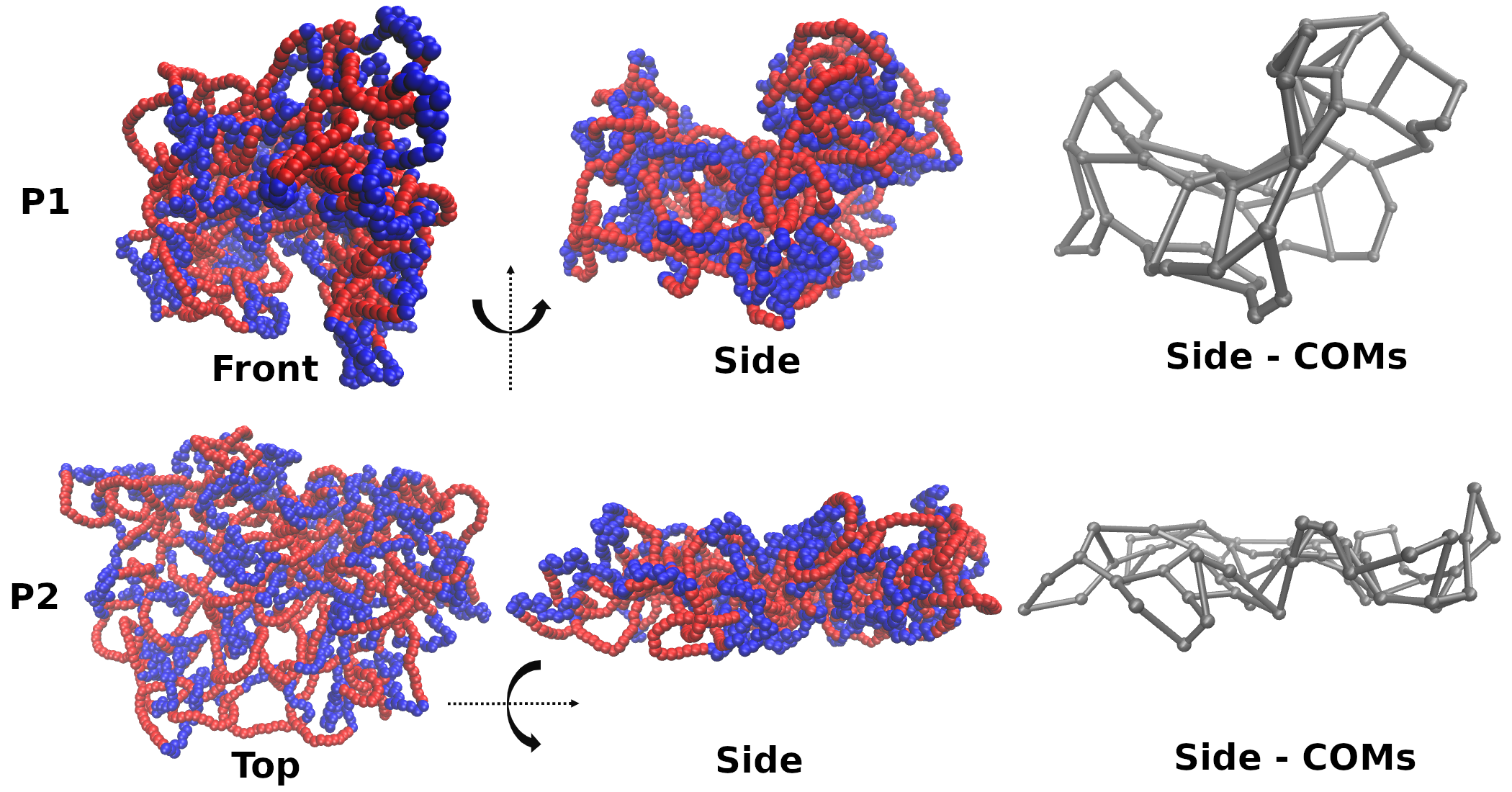}
 \caption{Typical equilibrium conformations of P1 and P2 chainmails with rings at 50\% rigid-segment coverage. The rigid and flexible blocks are colored in red and blue, respectively. The $P1$ chainmail is visibly bent in a saddle-like shape, while the $P2$ one is approximately flat. The rightmost panels are schematic representations of the same chainmails where the beads correspond to the rings' centers of mass (CoMs), and the bonds connect CoMs of linked  (neighboring) rings. The P1 and P2 conformations are not shown at the same scale.}
\label{fig:snapshots}
\end{figure*}

We considered all combinations of the selected ring compositions and linking patterns, ten in total, and used Langevin dynamics simulations to sample the equilibrated conformational ensembles. Typical configurations for P1 and P2 chainmails for rings with equally long rigid and flexible blocks (50\% rigid-segment coverage) are shown in Fig.~\ref{fig:snapshots}. For clarity, the front and side views of the chainmail are complemented with a wireframe representation, where the nodes represent the centers of mass of the rings, and the connecting edges represent the mechanical bonding of the corresponding rings.

Although they share the same ring composition and honeycomb lattice organization, the two representative chainmails present noticeable structural differences, which we next examined with a systematic quantitative analysis.

\subsection{Metric properties}

\begin{figure*}[ht!]
        \centering
\includegraphics[width=1.05\textwidth,height=0.65\textwidth,keepaspectratio]{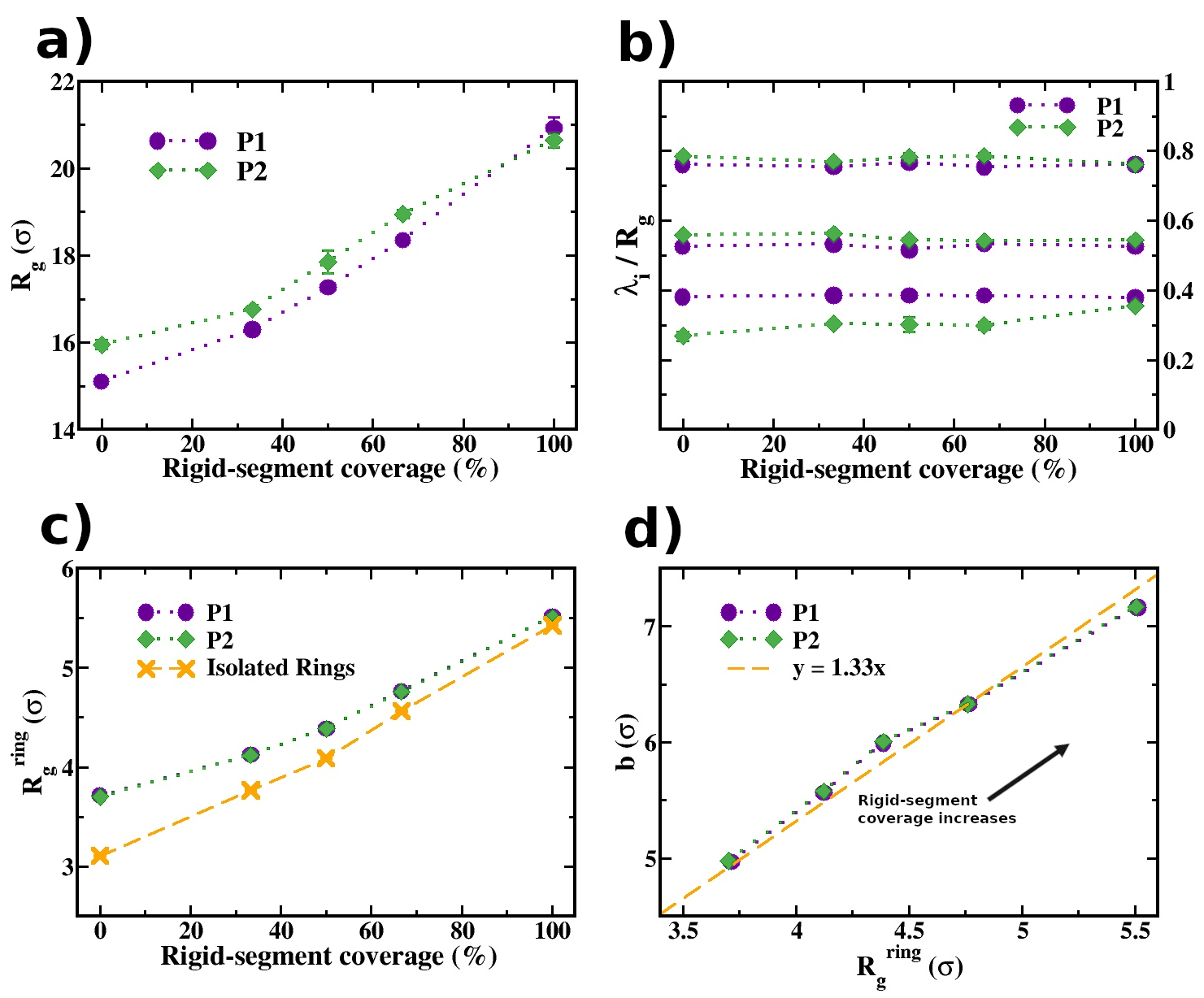}
        \caption{Metric properties of the system. a) Gyration radius of the chainmails ($R_{g}= \sqrt{<R_{g}^{2}>}$) at different compositions. b) Square root of the eigenvalues of the gyration tensor normalized by the $R_{g}$ at the corresponding composition ($\sqrt{<\lambda_{i}^{2}>/<R_{g}^{2}>}$). Normalization with $R_{g}$ evidences the scaling effect of the rigidity. c) $R_{g}^{\rm ring}=\sqrt{<(R_{g}^{\rm ring})^{2}>}$ of single rings inside chainmails with $P1$ (purple dots) and $P2$ (green diamonds). The orange dashed line corresponds to an isolated ring. d) Mechanical bond length, $b$, versus ring gyration radius, $R_{g}^{\rm ring}$, for P1 and P2 patterns. Error bars represent statistical uncertainty, calculated as half the difference between the average values of observables from the two collected trajectories, except for the $R_{g}$ of isolated rings, whose errors are computed with block analysis and bootstrapping.
        }
\label{fig:metric_properties}
\end{figure*}

Fig.~\ref{fig:metric_properties} presents a comparative overview of the average metric properties of chainmails with different compositions of the ring diblock copolymers and linking patterns.

The root mean square gyration radii ($R_{g} = \sqrt{<R_{g}^{2}>}$) of the chainmails are shown in panel $a$, where they are profiled for increasing rigid-segment coverage. The plot establishes that P2 chainmails are systematically larger than P1 ones. The size difference is maximum when the interlocked rings are fully flexible and progressively diminishes for increasing length of the rigid segment, becoming negligible in the limit case of fully rigid rings. These differences are a first indication of how the conformational space of mechanically bonded membranes with the same honeycomb connectivity and the same ring composition can be varied by solely intervening on the different modes for interlocking the same set of neighboring rings.

Further and more consequential differences emerge when comparing shape data, illustrated in Fig.~\ref{fig:metric_properties}b in terms of the ensemble-averaged eigenvalues of the gyration tensor. To discount effects related to the noted different overall sizes of P1 and P2, the three averaged eigenvalues were normalized to $R_g$. The first noteworthy feature is that the curves of the normalized eigenvalues are approximately horizontal, i.e.~independent of the rigid-segment coverage. Next, the flat profiles of the two largest eigenvalues of P1 and P2 are well-superposed and compatible within statistical errors. Significant deviations are instead observed for the smallest eigenvalue, which is systematically smaller for P2, indicating that it adopts flatter shapes than P1.
Similarly to the $R_g$ results, the difference is largest for fully flexible rings and negligible for rigid ones. We conclude that P2 is systematically flatter than P1.

The fact that the P1-P2 differences of both size and shape are largest for fully flexible rings is a noteworthy result in itself. In fact, one could have anticipated that the flexible rings, with their convoluted conformations, could screen better than rigid ones the anisotropic steric interactions resulting from the over- and under-passes of neighboring rings and hence be less conducive to emergent large-scale conformational properties.

A pertinent question related to the observations above is how exactly the conformations of the rings depend on being part of P1 or P2 membranes.
Analogously to the case of linear catenanes, this question is best addressed by comparing the size of the mechanically-bonded rings with isolated ones and by exploring the connection between the size of linked rings and their distance\cite{DehaghaniCatenaneExponent2020,LiDoubleScaling2021,LeiCatenaneShape2021,RauscherThermoMelt2020,chiarantoni2022rigidity} To this end, we considered the average mechanical bond length, defined as the average distance between the centers of mass of pairs of linked rings. For such an average, we considered all pairs except those with one or both rings at the boundary.

The data are presented in Fig.~\ref{fig:metric_properties}c,d and provide several indications. First, at all considered rigid-segment coverages, the size of chainmailed rings, $R_g^{\rm ring}$, and the mechanical bond length, $b$, are practically indistinguishable between P1 and P2. Second, chaimailed rings are significantly larger than isolated ones. This is consistent with previous results for one-dimensional (linear) catenanes, where the steric interactions of mechanically-bonded neighbors make concatenated rings larger than isolated ones\cite{chiarantoni2022rigidity}.
Finally, the two sizes $b$ and $R_g^{\rm ring}$ are proportional to each other.

The results reinforce the conclusion that the observed size difference of P1 and P2 chainmails directly arises from the different linking patterns, given that the size of the constitutive rings and mechanical bond length is the same for P1 and P2 at fixed ring composition. This fact implies that the two chainmail types cannot be distinguished by measuring simple metric properties locally, meaning at the scale of one or two rings.

The results of Fig.~\ref{fig:metric_properties} establish that the overall size and shape of the chainmails are emergent properties depending on their linking pattern and ring composition, as exemplified by the conformations in Fig.~\ref{fig:snapshots}. Specifically, (i) the rigid/flexible block-copolymer composition impacts primarily the overall size of the chainmails and only secondarily its shape, while (ii) varying the pattern of neighboring rings' linking modes at fixed chainmail connectivity has a limited but still discernible effect on size and very pronounced impact on the shape, which can be selectively varied from approximately flat to highly non-planar.

\subsection{Co-localization of rigid vs flexible regions of linked rings}

A relevant question for our system of linked block-copolymer rings is whether the flexible and rigid segments tend to segregate or mix in the chainmail due to their steric interaction at the mechanically bonded regions. This question arises naturally considering conventional systems of block copolymers, where effective repulsive interactions between segments of different types cause the latter to segregate, forming ordered structures such as ~cylindrical, lamellar, or spherical domains\cite{matsen1996origins,bates201750th,feng2017block}.

To address this point, for each sampled conformation, we first identified the contacting pairs of monomers in neighboring rings, that is, the pairs of monomers belonging to linked rings that are at a distance smaller than the interaction range of the WCA potential,  $\sqrt[6]{2}\sigma$. Next, each pair was assigned to one of three mutually exclusive classes, depending on whether the contacting monomers both belonged to rigid segments ($rr$), flexible segments ($ff$), or were in
segments of different stiffness ($fr$). By cumulating this information over the sampled conformations, we obtained the probabilities of the contacting pairs to be in the three classes, $P_{\rm rr}$, $P_{\rm ff}$ and  $P_{\rm fr}= 1 - P_{\rm rr}+P_{\rm ff}$.

\begin{figure*}[ht!]
        \centering
\includegraphics[width=1.05\textwidth,height=0.68\textwidth,keepaspectratio]{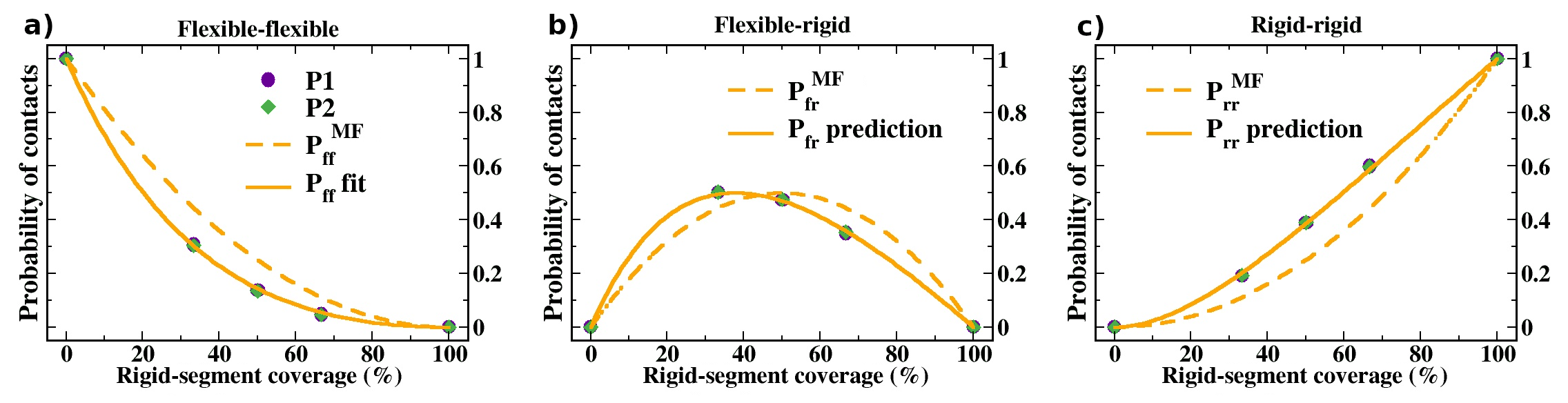}
\caption{Ring composition and contacting regions of linked rings. Probability of establishing a contact between two monomers belonging to flexible blocks (a), one belonging to a flexible and one belonging to a rigid block (b), and between two particles of rigid blocks (c). Results for P1 and P2 chainmails are shown with urple dots and green diamonds, respectively. Continuous lines correspond to the parameterized curve ($P_{\rm ff}$, panel a) and predictions ($P_{\rm fr}$ and $P_{\rm rr}$, panels b and c) of the model (Eq. \ref{eq:model} with $\Delta F \simeq 0.496K_{B}T$) while dashed lines show the predicted curves when there is no entropic gain/cost for the different combinations of contacts ($\Delta F = 0$). The error bars, calculated as half the difference between the average values of observables from the two collected trajectories, are smaller than the symbols.}
\label{fig:local_contacts}
\end{figure*}

The profiles of the three probabilities are given in Fig.~\ref{fig:local_contacts}a-c as a function of the rigid-segment coverage for both chainmail types.
The data show that the two limiting cases of membranes made of fully rigid ($P_{\rm rr}=1$) and fully flexible ($P_{\rm ff}=1$) rings are bridged by probability curves that are noticeably asymmetric with respect to the 50-50 composition. The data reveal that $rr$ contacts are over-represented compared to the baselines obtained by mean-field-like (MF) combinatorial considerations based on the fraction of monomers in rigid segments, $x$, which yield  $P_{\rm ff}^{\rm MF} =(1-x)^2$, $P_{\rm fr}^{\rm MF} =2x(1-x)$, and $P_{\rm rr}^{\rm MF} =x^2$, see dashed curves in Fig.~\ref{fig:local_contacts}.

This bias is evident when considering the 50-50 composition where the probability of $rr$ contacts ($P_{\rm rr}(50\%)\simeq 0.39$) is not equal to the value obtained in the flexible case ($P_{\rm ff}(50\%)\simeq 0.14$) as in the mean-field estimate for the balanced case, but is approximately threefold larger.
Furthermore, when as much as one-third of the rings' contour is flexible, the probability of $ff$ contacts is only marginally higher than in the fully rigid ring case.

The robustness of the observed segregation tendency of $rr$ contacts was tested by adopting a different criterion for selecting contacting monomer pairs. Specifically, for any two mechanically bonded rings, we exclusively selected the two monomers (again, one per ring) at the shortest distance. If the latter was smaller than the contact cutoff distance, $\sqrt[6]{2}\sigma$, then the pair was assigned to the $ff$, $rr$ or $rf$ class as before. Otherwise, it was labeled as not in contact. This alternative counting of contact interactions is an apt complement of the one discussed before because a single pair of monomers is considered for any two linked rings. The data, shown in  Fig. S6 of the Supplementary Material, present a bias for $rr$ contacts analogous to that of Fig.~\ref{fig:local_contacts}. An interesting insight emerging from the alternative analysis is that, although the rings in P1 and P2 have about the same average size and mechanical bonding distance, the fraction of selected pairs not in contact is larger for P1 chainmails.

Based on these converging results and considering that non-bonded monomers exclusively interact via a steric repulsion, we conclude that contacts between flexible regions of mechanically bonded rings are entropically disfavoured. This is arguably due to linked rings having more wiggle room or conformational freedom when they come into contact with their smooth, rigid regions rather than the crumpled, flexible ones.

To estimate the entropic cost of bringing flexible regions in contact, we fitted the probability data with a minimalistic model that improves on the mean-field-like combinatorial argument by introducing an excess free energy, $\Delta F$, for each contacting monomer belonging to a flexible segment rather than a rigid one,
\begin{eqnarray}
        P_{\rm rr}(x) &=& \frac{x^{2}}{\mathcal{N}}, \nonumber \\
        P_{\rm fr}(x) &=& \frac{2x(1-x) e^{-\Delta F/k_BT}}{\mathcal{N}}, \nonumber \\
        P_{\rm ff}(x) &=& \frac{(1-x)^{2} e^{-2\Delta F/k_BT}}{\mathcal{N}}
\label{eq:model}
\end{eqnarray}
\noindent where $\mathcal{N}=[x + (1-x) e^{-\Delta F/k_BT}]^2$ is the normalization factor.

Although the model of Eq.~\ref{eq:model} is crude, it can reproduce and predict the observed data remarkably well. To show this, we fixed the single free parameter by solely fitting the $P_{\rm ff}$ data (solid line in Fig.~\ref{fig:local_contacts}a), obtaining $\Delta F \simeq 0.496 k_{B}T$.
Next, we used the fixed $\Delta F$ parameter to obtain predictions for $P_{\rm fr}$ and $P_{\rm rr}$. The resulting curves agree well with the observed contact probabilities, see solid curves in panels b and c, respectively.

The inferred $\Delta F$ value is comparable to the thermal energy, which is not surprising considering the entropic origin of the observed bias for $rr$ contacts. At the same time, because the contour length of the regions establishing the interlockings does not scale linearly with the ring's contour length\cite{caraglio2017physical,Caraglio_et_al_polymers_2017}, we expect that the magnitude of the $rr$ entropic segregation could significantly vary with the length of the rings, which could thus be a relevant design parameter for tuning chainmail properties.

\subsection{Gaussian curvature}

Taking into account the qualitatively different shapes of the representative P1 and P2 conformers in Fig.~\ref{fig:snapshots}, we systematically analyzed the Gaussian curvature, ${K}_G$, of P1 and P2  chainmails both locally and globally. A further motivation for this analysis is to compare our honeycomb chainmails with previous results based on different architectures\cite{polson2021}. For such systems, it was consistently observed that
the membranes invariably acquired a cup-like shape, i.e. a spontaneous positive Gaussian curvature, independently of the lattice shape or
the thickness of the rigid constitutive rings\cite{polson2021}.

\begin{figure}[ht!]
        \centering
\includegraphics[width=0.5\textwidth,height=0.5\textwidth,keepaspectratio]{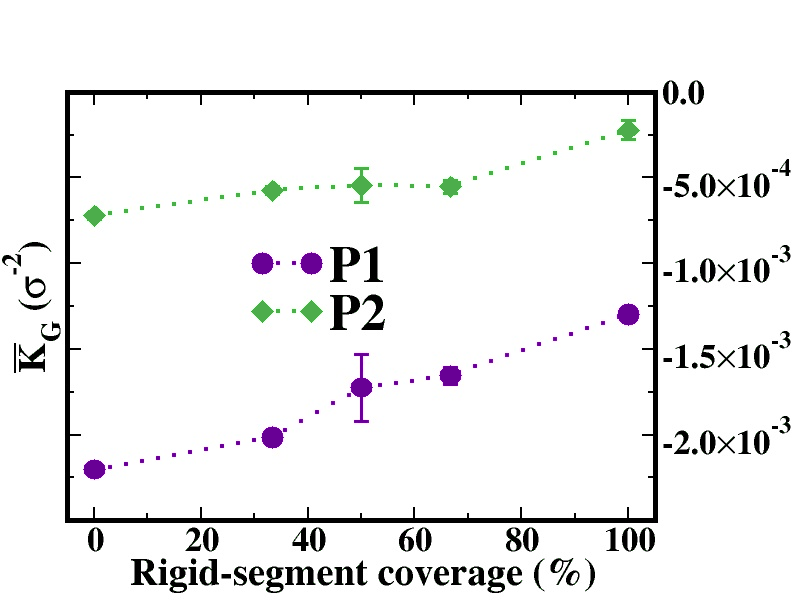}
        \caption{Average Gaussian curvature of the membranes for both linking patterns. P2 membranes have a $|\overline{K}_{G}|$ an order of magnitude smaller than $P1$ membranes. In both cases $|\overline{K}_{G}|$ decreases with the rigidity. Error bars represent statistical uncertainty, calculated as half the difference between the average values of observables from the two collected trajectories.}
\label{fig:kgavr}
\end{figure}

Fig.~\ref{fig:kgavr} presents the average global Gaussian curvature, $\overline{K}_G$, for various ring compositions of the two types of membranes.
The plot reveals that both P1 and P2 have a negative spontaneous curvature.
At any given composition, the two membrane types are clearly distinguished by the magnitude of $\overline{K}_G$,  whose modulus is at least fivefold larger for P1 than for P2.
Because negative Gaussian curvatures are associated with saddle-like shapes, one has that saddles such as the one of Fig.~\ref{fig:snapshots} are indeed typical for P1, and that P2 conformers, while appearing overall flat, are slightly concave-convex, too.
In addition,  $\overline{K}_G$ becomes significantly more negative as the length of the flexible segment increases. A two-fold increase of $|\overline{K}_G|$ for P1 is observed going from fully rigid to fully flexible rings, a point further discussed later.

 The observed negative curvature is remarkable considering that, to our knowledge, only positive Gaussian curvatures have been reported before for chainmails, constructed starting from square and triangular architectures\cite{polson2021}. Here, the systematic emergence of negative $\overline{K}_G$ -- irrespective of the linking mode (P1/P2) or ring composition -- gives a strong indication that the honeycomb lattice connectivity is conducive to saddle-like shapes. This predisposition can be very strongly modulated by the modes that neighboring rings are linked with one another and, to a comparable but lesser extent, by the ring composition.

For a more detailed insight into the emergence and persistence of saddle-shaped conformations, we analyzed $K_G$ at the local level. To do so, we considered the longest trajectory and, for each sampled conformation, we first computed the instantaneous value of $K_G$ in the neighborhood of each ring. Then we calculated the time-averaged value of the local curvatures, and reported them as heatmaps on a regular (flattened) representation of the underlying honeycomb lattice.

\begin{figure*}[ht!]
        \centering
\includegraphics[width=0.92\textwidth,height=0.88\textwidth,keepaspectratio]{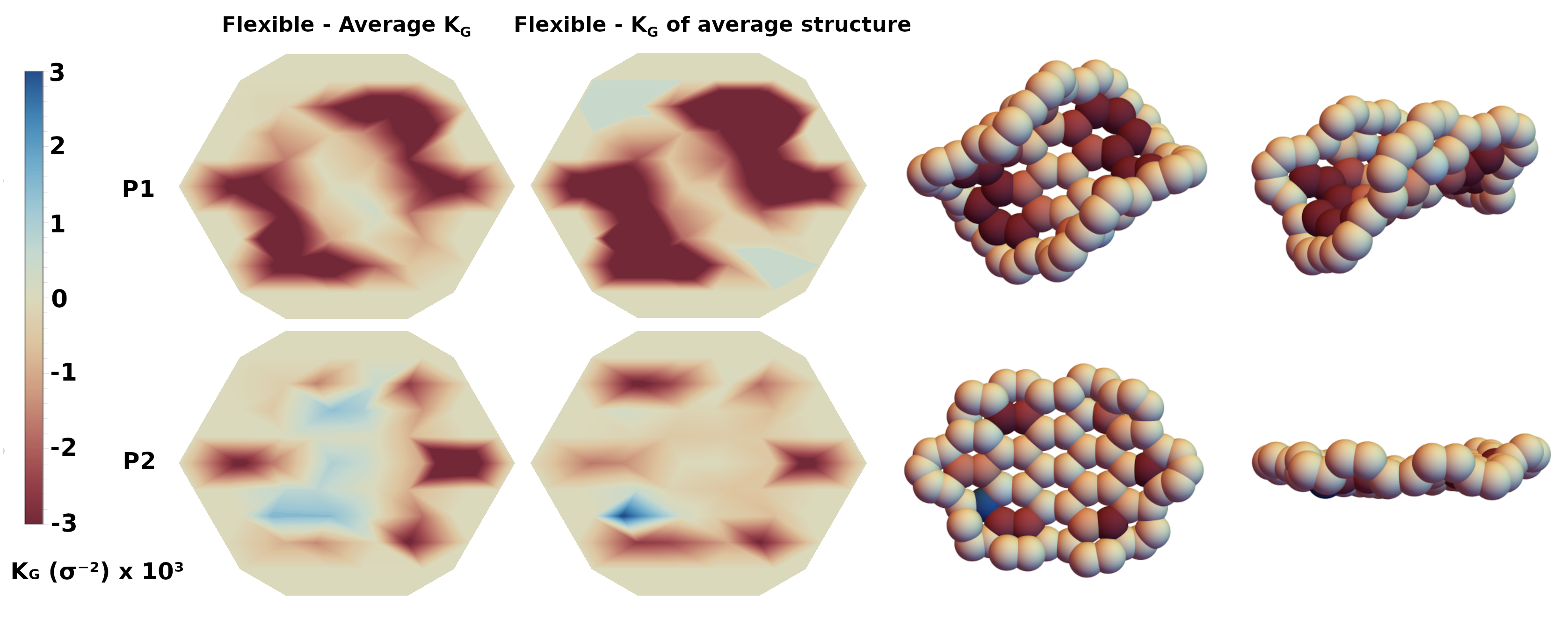}
        \caption{Local Gaussian curvature, $K_{G}$, computed for the $2.5\times 10^{5}\tau$ trajectories of P1 (top row) and P2 (bottom row) chainmails with fully-flexible rings. The leftmost column shows the heatmaps of the local values of $K_{G}$ averaged over the trajectories. The central column shows the $K_G$ heatmap of the average P1 and P2 structures. Top and side views of the latter are shown on the right, each bead representing the center of mass of one ring. The beads are color-coded based on the corresponding local $K_{G}$ values.
        }
        \label{fig:gaussian_curvature1}
\end{figure*}

The resulting heatmaps for the case of fully flexible rings, where $|K_G|$ is largest, are shown in the leftmost column of Fig.~\ref{fig:gaussian_curvature1}.
The time-averaged heatmap of P1 is biased towards negative $K_G$ values, characteristic of saddle-like states. On the contrary, the time-averaged heatmap of the P2 membrane of fully flexible rings displays no significant negative curvature pattern but smaller domains with both signs of $K_G$.

We repeated the curvature analysis for the average chainmail structure to better interpret the time-averaged heatmaps. This structure was obtained with a structural alignment of all configurations (snapshots) sampled in the trajectory. To this end, we employed the Kabsch optimal roto-translation\cite{kabsch1978discussion} to align the centers of mass of each conformer with those of a reference one. The reference configuration was selected as the conformer with the smallest average root-mean-square deviation from all other snapshots in the trajectory.
The resulting P1 and P2 average structures are shown in the right panel of Fig.~\ref{fig:gaussian_curvature1}, where the beads correspond to the centers of mass of the rings and are colored according to their local $K_G$ values, using the same color scheme of the accompanying heatmaps.

On the one hand, the average structures provide an intuitively interpretable counterpart of the $K_G$ heatmaps, evidencing the negative $K_G$ regions of the P1 saddle and the localized $K_G$ domains of the slight P2 undulations.
On the other hand, the fact that the heatmaps of the time-averaged curvatures closely resemble those of the $K_G$ computed for the average structure implies that the observed curvature patterns are very persistent in time and correlated over timescales exceeding the duration of the simulations. We recall that the latter was chosen, as customary, to largely exceed the autocorrelation time of the gyration radius (Fig. S2 in the Supplementary Material), which is thus amply surpassed by the long lifetimes of the observed curvature patterns. The latter eventually ought to switch between the different states compatible with the chainmail symmetry; consistent with this, symmetry-related patterns are observed in different trajectories (Fig. S8 in the Supplementary Material).

\subsection{Ring composition: effect on $K_G$ magnitude and istotropic rescaling}

\begin{figure*}[ht!]
        \centering
\includegraphics[width=0.92\textwidth,height=0.88\textwidth,keepaspectratio]{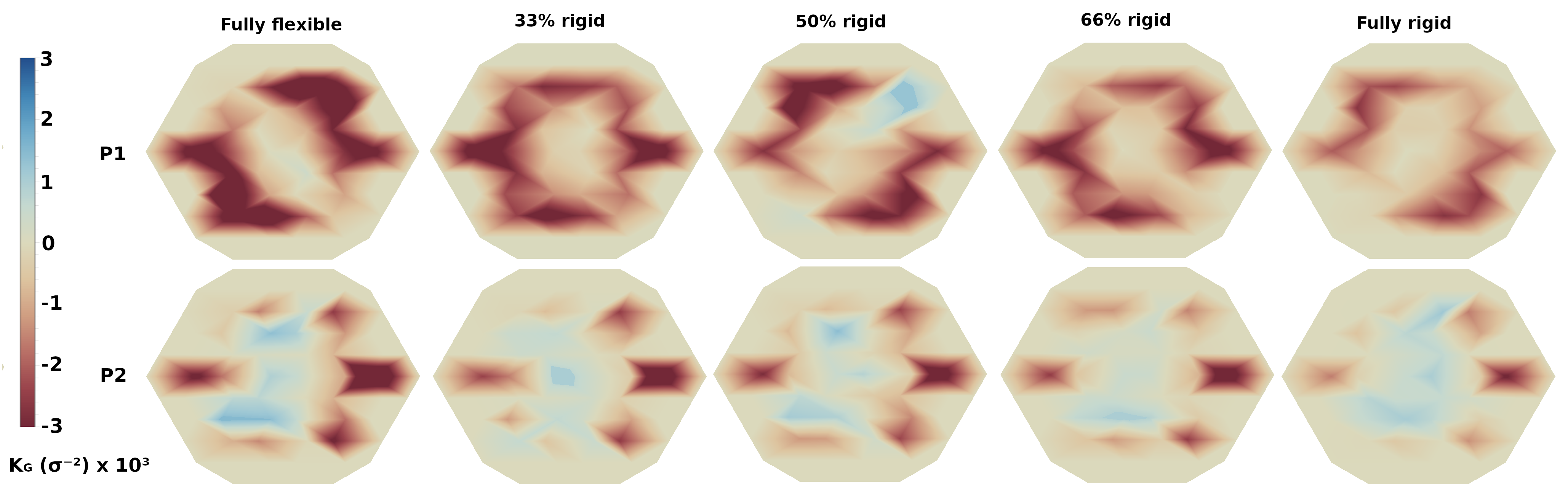}
        \caption{Heatmaps of the local values of $K_{G}$ averaged over the $2.5\times 10^{5}\tau$ trajectories for P1 (top row) and P2 (bottomo row) chaimails for the indicated ring compositions. The magnitude of $K_G$ reduces with the increasing size of the rigid segment.
        }
        \label{fig:gaussian_curvature2}
\end{figure*}

The spontaneous curvature results discussed in Fig.~\ref{fig:gaussian_curvature1} for membranes with fully flexible rings are consistently observed across the considered ring compositions, albeit to a diminishing degree as the rigid-segment coverage grows larger. This conclusion emerges from the series of heatmaps in Fig.~\ref{fig:gaussian_curvature2}. The images also illustrate the point noted above that the conformational ensemble of the chainmails comprises states related by the discrete symmetries of the chainmails, which depend on the symmetry of the connectivity network, the pattern of linking modes,  and the shape of the finite chainmails. Indeed, the principal axes of the various P1 heatmaps can be aligned along one of three possible symmetry axes.

\begin{figure*}[ht!]
        \centering
\includegraphics[width=1.05\textwidth,height=0.65\textwidth,keepaspectratio]{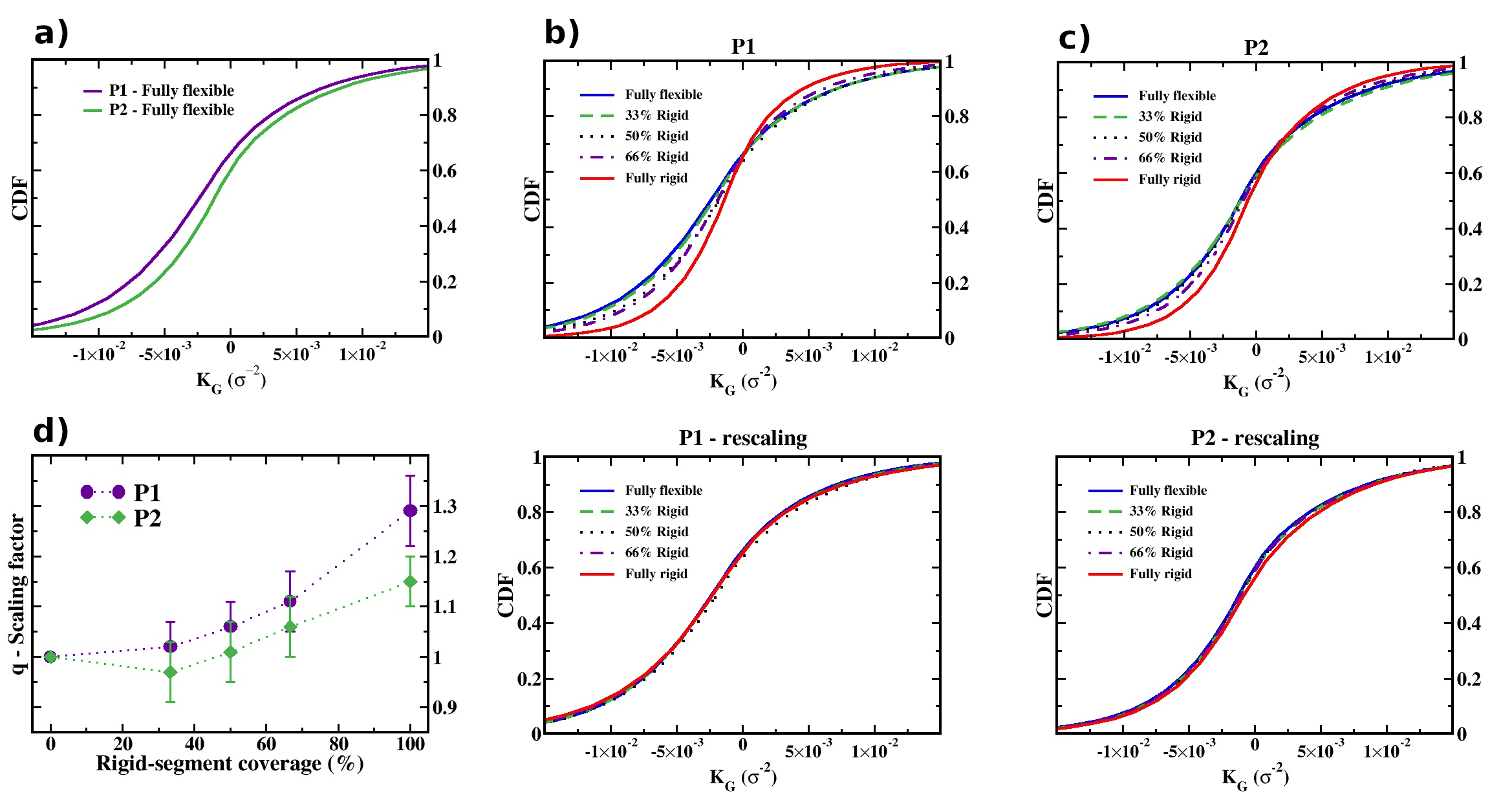}
        \caption{Cumulative distribution function (CDF) of the local Gaussian curvature $K_G$. a) CDFs of P1 and P2 membranes with fully flexible rings, computed for the $2.5\times 10^{5}\tau$ trajectories. The effect of varying the rigid-segment coverage for the CDF of P1 and P2 chainmails are respectively shown in the top b) and c) panels. The bottom panels show that the same CDF curves can be superposed with a suitable isotropic length rescaling of chainmails. The optimal isotropic scaling factors yielding the shown superpositions are given in panel d). The error bars span the values of $q$ that provide a quadratic error under $5 \%$ relative to the peak of the reference distribution.
        }
\label{fig:scaling_arguments}
\end{figure*}

The combined effect of ring composition and ring linking modes on the Gaussian curvature is recapitulated in Fig.~\ref{fig:scaling_arguments}, which presents the cumulative distribution function (CDF) of the local curvatures of P1 and P2 chainmails sampled across an entire trajectory.
The P1 and P2 CDFs of panel (a) correspond to the case of fully flexible rings. The P1 curve is shifted to the left, i.e., skewed towards more negative values, consistently with the larger value of the average curvature reported above.
Instead, the effect of ring composition is shown in the top panels of Fig.~\ref{fig:scaling_arguments}b,c. For both P1 and P2, the sigmoidal CDF curves cross approximately at the same point, although with different slopes, indicative of sharper $K_G$ distributions for higher rigid-segment coverage.
The median $K_G$ values, corresponding to CDF equal to 0.5, are shifted towards more negative values as the rigid-segment coverage diminishes, again consistent with the previous conclusions based on Fig.~\ref{fig:kgavr}.

\subsubsection{Ring composition and isotropic size rescaling}

As we noted in connection to Fig.~4b, the normalized eigenvalues of the average gyration tensor of P1 appear to be constant across the considered ring compositions. The same property holds for the top two eigenvalues of P2, with the third, smallest one, varying by no more than 30\%.
These results suggest that the conformational ensemble of the chainmails is largely self-similar, meaning that the P1 or P2 average gyration tensors for different compositions differ primarily by an isotropic scale factor.

To more directly ascertain the size rescaling effect, we investigated whether the probability distributions of the local Gaussian curvature at different rigid-segment coverage are consistent with a simple isotropic size rescaling of the membranes.
To this end, we consider the affine transformation of a membrane corresponding to multiplying the Cartesian coordinates of all the nodes by the same scale factor, $q$. Based on the Gaussian curvature definition, the local Gaussian curvatures of corresponding points on the original and isotropically rescaled membrane satisfy
$\tilde{K}_G/K_G =q^{-2}$, where the $\sim$ superscript refers to the transformed membrane.

Accordingly, we took for reference the $K_G$ cumulative distribution of the fully flexible rings case and asked whether the CDFs for any other rigid-segment coverage could be collapsed on it by rescaling the argument, i.e., the $x$ axis of the plots in Fig.~\ref{fig:scaling_arguments}b,c by a suitable multiplicative factor, $q^{-2}$, corresponding to rescaling by $q$ the coordinates of the conformational ensemble of the membranes.

Using a best-fit procedure to superpose the CDFs yields the curves in the bottom panels b and c for types P1 and P2, respectively. The collapse of the curves for all ring compositions is noticeably good, particularly for the P1 chainmail type.

It is interesting to consider the real-space scaling factors, $q$, inferred from the best fit. The data, given in Fig.~\ref{fig:scaling_arguments}d, indicate that $q$ has a general increasing trend with the length of the rigid block. Pleasingly, the reported values of $q$ are comparable with the ratios of other characteristic lengthscales, such as the chainmail gyration radius and mechanical bond length. For instance, the $CDF[K_G]$ curves for fully rigid and fully-flexible rings are optimally superposed for $q$ equal to $1.26$ (P1) and $1.14$ (P2), which are not dissimilar from the corresponding ratios of $R_G$, 1.39 (P1) and 1.29 (P2), and of $b$, 1.44 (both P1 and P2) in Fig.~\ref{fig:metric_properties}.

Overall, the above results thus support the notion that the composition of the rings primarily defines the overall size of the chainmails, not their shape, which can dramatically vary with the pattern of ring linking modes.

%% file: sections/conclusions.tex
\section{Conclusions}

We considered two-dimensional chainmails of interlocked ring polymers and used Langevin dynamics simulations to study the effects of ring composition and chainmail topology on the equilibrium conformations.
We considered two chainmail types, P1 and P2, with the same honeycomb overall architecture but different local patterns of over- and under-crossings of the linked rings; the latter were modeled as diblock copolymers made of rigid and flexible segments at five different relative compositions.

By examining various local and global metric observables, we found that ring composition and linking patterns affect chainmails in different and complementary ways. Specifically, while the former primarily sets the overall size of the chainmail, the latter defines the shape.

The P1 chainmail, where all rings have the same linking pattern, innately adopts saddle-like shapes characterized by negative Gaussian curvatures both locally and globally. Instead, the P2 chainmail, where the above pattern is interrupted by regularly spaced rows of differently linked rings, is flat. The imprinting of the linking topology is so strong that the above shapes emerge systematically throughout the considered range of block copolymer compositions, from fully rigid to fully flexible.

In line with this result, we found that the conformational ensembles of chainmails with the same linking pattern but different ring compositions can be superposed with an affine transformation corresponding to a uniform rescaling. The composition-dependent scaling factor is defined by the characteristic gyration radius of the individual rings or, equivalently, by the mechanical bond length.

While it is intuitively plausible that varying ring composition can modulate the chainmail's overall size via the linked rings' size and distances, the shape equivalence under isotropic rescaling is a noteworthy and non-intuitive result. For instance, because flexible rings have a smaller metric footprint and are more convoluted than rigid ones, one could have surmised that the interface regions of linked flexible rings would be intricate, preventing the details of the linking pattern from reverberating globally. It is thus surprising that such topological screening is not observed and that average structures of chainmails with fully rigid and fully flexible rings can be superposed by isotropic rescaling.
At the same time, ring composition does have implications for the microscopic organization of the chainmails and precisely for type of blocks preferentially co-opted at the interlocked regions. In fact, we observed that rigid blocks are systematically over-represented over flexible ones.

The above results have several implications worthy of future investigations.
First, they establish that at least one linking pattern can yield conformations with negative Gaussian curvature. This class of shapes, which has not been reported before for chainmails, is relevant in supramolecular synthetic chemistry\cite{woods2022shape,woods2023saddles}. Our results show that by extending considerations to metamaterials based on mechanical bonding, the saddle shape can be achieved with an entirely new design principle, namely by controlling the local linking patterns.
Second, the observed robustness of the shape-conditioning effect of the linking pattern suggests that chainmail architectures other than the honeycomb one could yield saddle-shaped membranes, too. Accordingly, it would be relevant to verify this hypothesis. Next, the shape equivalence - up to an overall scale factor - of chainmails of different ring compositions indicates that the latter could be a relevant tunable parameter for designing chainmails according to given size specifications. Additionally, It would be interesting to explore the effect of varying the length of the constitutive rings to ascertain if it could have global implications beyond the expected impact on chainmail size. A related point would be to extend considerations to chainmail patches of increasing size\cite{polson2021}. Although the latter point is beyond the scope of the present study, preliminary results presented in Fig.~S10 in the Supplementary Material suggest that the shape equivalence under isotropic rescaling might hold for chainmails with different sizes, too. Finally, it would be interesting to co-opt analysis methods developed for networks of linked polymers\cite{grillet2012polymer,doi:10.1021/acs.macromol.1c00176,katashima2021rheological} to establish the elastic properties of chainmails of different ring composition and linking patterns.

In conclusion, we have introduced a set of topological and composition features that lead to a rich phenomenology of membrane conformations at different scales, where simple rationalizing rules emerge despite the complexity of the system. Given the wide possibilities in which the basic ingredients can be combined, we expect many novel scenarios to be uncovered upon systematically exploring the physics of this setup, advancing the rational design of mechanically bonded meta-materials.